\documentclass{article}
\usepackage{amsmath}
\usepackage{graphicx}
\usepackage{xcolor}
\usepackage{appendix}
\usepackage{latexsym, amsfonts, amssymb, txfonts, pxfonts}
\usepackage{jheppub}

\title{Revisiting zero modes and cluster decomposition at the late-time boundary of de Sitter}
\author[a,c]{Murat Önem}
\author[a,b]{Gizem \c{S}eng\"or}
\affiliation[a]{Physics Department, Boğaziçi University 34342 Bebek, İstanbul Turkey}
\affiliation[b]{Feza Gürsey Center  for Physics and Mathematics, Boğaziçi University, Istanbul, Turkey}
\affiliation[c]{Department of Physics, North Carolina State University, Raleigh, NC 27607, USA}
\emailAdd{gizem.sengor@bogazici.edu.tr, monem@ncsu.edu }
\abstract{ We revisit the literature on locality on de Sitter with the goal to organize the main results with respect to the representation theory of the isometry group of four dimensional de Sitter. We make use of the late-time behavior of two-point functions of principal and discrete series representation, both in physical and in field space and compare the role of the zero modes. Our overall conclusion is that when it comes to locality on de Sitter, analyzed in terms of cluster decomposition, the principal series representation that capture matter fields and discrete series representations that capture gauge fields show different behavior. Focusing on scalars as a first analysis, matter fields show explicit signs of respecting cluster decomposition while gauge fields do not.}

\begin{document}
\maketitle

\section{Introduction}
\label{sec:intro}

From the perspective of the Lagrangian, a free massless scalar field, on any background, looks simpler compared to its massive counter-part because one only has the kinetic term to deal with. Yet, from the perspective of representation theory, free heavy scalars on a rigid de Sitter background are more straightforward to handle than the massless one.  One such example shows up in recognizing scalar states at the late-time boundary of de Sitter. Out of such states that realize unitary irreducible representations of the isometry group, it is straightforward to normalize those that arise from heavy fields \cite{unitarity}, and correspond to \emph{principal series} representations, while subtleties arise in the case of the massless scalar \cite{Sengor:twopoint} which is akine to \emph{discrete series} representations. With this motivation, the goal of this work is to revisit results on cluster decomposition on de Sitter with a focus on the treatment of zero-modes with respect to representation theoretic categories. 

Cluster decomposition address the locality of a given quantum field theory. In a local quantum field theory the expectation is that experiments spatially set apart at large enough distances do not affect each other. At a practical and calculable level for quantum fields on flat space, this implies that the S-matrix element for the overall processes factorizes into S-matrix elements of near by processes, as given in equation (4.3.1) of \cite{Weinberg:1995mt}. The position space statement of the cluster decomposition principle is that the connected part of the S-matrix vanishes when it involves states that are very far apart (in different clusters). In momentum space this determines how smooth the connected part of the S-matrix can be, that is various poles and branch-cuts at certain values of the momentum are allowed but there shouldn't be singularities as severe as delta functions. For de Sitter, S-matrix formulation is not readily available due to lack of global time translation invariance and the observable quantities are equal time two-point functions, where the representation theory shows itself best in the late-time limit.

The main focus of the literature on locality in de Sitter has been on massless scalar fields \cite{Allen:1985ux,Anninos:2011kh} for a long time. There are four categories of unitary representations of the de Sitter group, \emph{principal series, complementary series exceptional series} and in even number of spacetime dimensions \emph{discrete series}. Principal, complementary and discrete series are irreducible as well as unitary, exceptional series on the other hand reduce into discrete series in dimensions where the later category exists. Massless scalars are one of the members of the exceptional series representations that share similar properties with discrete series category where the discrete series are considered to correspond to the cases with nonzero spin by most authors. The discrete series category is expected to capture gauge fields. Therefore most of the effort towards locality on de Sitter in literature focuses on one of the members of the discrete series representations. This particular representation seems to disregard locality in a complicated way. In physical space, the signature of locality on de Sitter is expected to show up in the late-time behavior of position space two-point functions. Earlier works indicate differences to the Wightman function \cite{Bros_2010, anninos2023discreet} between the principal and discrete series representations and logarithmic growth for the later category \cite{Benincasa:2022gtd}. In field space, one can study the probability distribution of distance between two field values which has been interpreted to exhibit ultrametricity in the case of a massless scalar \cite{Anninos:2011kh}. The field space analysis of \cite{Anninos:2011kh} have been extended to scalars of general mass in \cite{Roberts:2012jw} .

In this work we revisit these two concepts related to locality: cluster decomposition in position space and field space with a focus on the treatment of zero-modes and categorization with respect to representation theory. We consider the late-time limit both for ease of calculations because it makes the identification of the representations easier as shown in \cite{unitarity,Sengor:twopoint,Sengor:particles} and because it can be interpreted as the thermodynamic limit from a statistical point of view.

In section \ref{sec:zeromodes} we will see that the two-point function of a scalar field on $S^4$ signals a problem with the zero modes for discrete series representations alone and in section \ref{sec:clust dec in position space} the two-point function of a scalar field on $dS_4$ violates cluster decomposition in position space, only in the massless case. 

At first sight this observation suggests that the violation of cluster decomposition in position space that appears only for the discrete series representations may be due to a mishandling of the zero-modes for this category. In section \ref{sec:clust dec in position space} we will see that the problem of locality for discrete series is more subtle then that by carrying out an analysis on removing the zero-modes.

We will keep visiting the issue with the zero-modes in section \ref{sec:cluster decomposition in field space}, where we discuss signs of cluster decomposition from a field space perspective, to highlight how zero-modes can be problematic even beyond the discrete series representations. 

To set notation and some background definitions we give a short summary of de Sitter representation theory in section \ref{sec:Summary of Scalar de Sitter representations} and the methods we will make use of in calculating two point functions in section \ref{sec:Review of two-point functions on the $S^4$ and $dS_4$}, where we also introduce the concept of zero-modes. We summarize our results in section \ref{sec:Conclusions and outlook} and give an outlook for future discussions.

\section{Summary of Scalar de Sitter representations}
\label{sec:Summary of Scalar de Sitter representations}
In $d+1$ dimensions, the isometry group of de Sitter spacetime, $dS_{d+1}$, is the group $SO(d+1,1)$. This group, which we will refer to as the \emph{de Sitter group}, is also the conformal group of $d$ dimensional Euclidean space. Since the detailed works of Harish-Chandra \cite{herb1991} the representation theory of the de Sitter group is well known. It is of relevance both in the context of Euclidean Conformal Field Theory, where \cite{Dobrev:1977qv} gives a detailed summary on bosonic representations and their applications, and in the context of de Sitter Quantum Field Theory. Here we focus on the de Sitter representations from the aspect of de Sitter Quantum Field Theory, and limit our discussion only to the scalar case. 

In the applications of de Sitter representations in the context of quantum field theory, the unitarity of these representations have played the key role in literature. Some of the highlights from this perspective include the identification of a forbidden mass range for spin-2 fields on de Sitter \cite{Higuchi:1986py}, unitarity of scalar tachyons on de Sitter \cite{Bros_2010} which provide a starting point for the study of gauge fields \cite{anninos2023discreet}, appearance of bosonic unitary irreducible representations at the late-time boundary of de Sitter \cite{unitarity} and  their contribution to cosmological correlators \cite{Di_Pietro_2022,Isono:2020qew,Sengor:twopoint} as well as the unitarity of fermionic representations \cite{Pethybridge:2021rwf,Letsios:2023qzq,Letsios:2025pqo,Higuchi:2025pbc} to list a few. Here we will briefly review the categorization of scalar representations as they appear in the quantization of a scalar field on a rigid de Sitter background, along the lines of \cite{Sengor:particles}.

Focusing on $dS_4$, the group $SO(4,1)$ is the Lorentz group for five dimensional Minkowski. From that perspective it only contains boosts and rotations on a five dimensional Minkowski spacetime. In terms of the four dimensional de Sitter coordinates, these five dimensional Minkowksi boosts and rotations appear as three spatial rotations, which make up the $SO(3)$ subgroup, a single dilatation that is the $SO(1,1)$ subgroup, three Special Conformal Transformations, each of which is parametrized along each spatial direction, and three spatial translations also parametrized along each spatial direction. Contrary to four dimensional Minkowski where the isometry group is the Poincaré group $ISO(3,1)$, the isometries of de Sitter do not involve time-translations and this makes all the difference between these two Lorentz groups. 

To discuss the representation theory, it is helpful to first discuss how to label representations. We can collect this information from the quadratic Casimir of the corresponding group algebra and the Cartan subgroup. The quadratic Casimir commutes with all the generators of the algebra and as such its eigenvectors are useful to set label a basis of eigenstates. The quadratic Casimir eigenvalues of the de Sitter algebra depend on the eigenvalues of quadratic Casimir for the spatial rotation generator, which we will denote as $l$ and refer to as spin, and the dilatation generator, which is the scaling dimension, denoted here as $\Delta$. Another clue towards how to discuss the representations lies in the Cartan Subgroup, which gives the maximum number of simultaneously diagonalizable subgroups. For the de Sitter group, the Cartan Subgroup is made up of dilatations and rotations. Thus, judging by the quadratic casimir eigenvalues, the representations are best labeled by spin and scaling dimension and states labeled as such will be the eigenstates of dilatation and rotation subgroups simultaneously.

The scaling dimension $\Delta$ is associated to how operators transform under dilatation, which is a constant rescaling of the coordinates. For the conformal planar patch de Sitter metric 
\begin{align}
    \label{metric} ds^2=\frac{-d\eta^2+d\vec{x}^2}{H^2|\eta|^2},
\end{align}
where $\eta\in(-\infty,0)$ is the conformal time and $H=l^{-1}_{dS}$ is the Hubble parameter which in the case of de Sitter is constant and sets the de Sitter length scale, the dilatation transformation is
\begin{align}\label{eqn:dilatation}
    \eta\to\lambda \eta,~~x^i\to\lambda x^i~~\text{where}~i=1,2,...,d.
\end{align}
Under \eqref{eqn:dilatation} an operator that transforms as
\begin{align}
    \mathcal{O}(\lambda\vec{x})=\lambda^{-\Delta}\mathcal{O}(\vec{x})
\end{align}
is said to have scaling dimension $\Delta$. For the de Sitter group, the scaling dimension has a specific dependence on the number of spatial dimensions and mass of the field. The scaling dimension is
\begin{align}
    \Delta=\frac{d}{2}+c
\end{align}
where $c$ is called the scaling weight and for unitary representations it is allowed to be real or purely imaginary. The quadratic casimir eigenvalue is
\begin{align}\label{eqn:so(d+1,1)casimir eigenval}
    c_2&=l(l+d-2)+c^2-\frac{d^2}{4}\\
    &=\Delta(\Delta-d)+l(l+d-2),
\end{align}
which is real for $c$ real and purely imaginary.

Depending on the scaling weight $c$ being purely imaginary or real, the unitary irreducible representations fall under different categories. Unitarity implies having a well defined inner product on specific function spaces which remains invariant under the action of the unitary representations. The well defined inner product is straight forward when $c$ is purely imaginary and this defines the first category: \emph{principal series}. When $c$ is real, the well defined inner product requires the action of intertwining operators which generate a shadow transformation. The shadow transformation is a similarity transformation that maps the scaling dimension in the specific way such that $\Delta\to\tilde{\Delta}=d-\Delta$ while keeping $l$ invariant. The normalizability and invertibility of the intertwining operator split the case of real $c$ into further categories. 

All together, the unitary irreducible representations of the de Sitter group fall under three main categories: \emph{principal series, complementary series, exceptional series}. For unitary irreducible representations realized by symmetric traceless tensors, the exceptional series category further breaks into \emph{discrete series representations} at the specific values of $d$ equals to $1$ and $3$ . The exceptional series representations themselves are composed of four subcategories. One of these capture scalars and another one captures fields with nonzero spin. While the terminology "discrete series" is commonly reserved for fields with nontrivial spin, the concerns regarding unitarity, as understood in the form of well defined inner products on functions that are elements of specific function spaces on which the unitary irreducible representation act, work similarly among the four categories of exceptional series. Therefore we will refer to the scalar unitary irreducible representation that  belongs to the exceptional series type-I at $d=3$ also as the discrete series representation. Reference \cite{unitarity} discuss the principal and complementary series inner product with explicit examples from the late-time behavior of scalar fields while \cite{Sengor:twopoint} carries on the discussion for discrete series.

Focusing on scalars in $d=3$, the unitary irreducible representation categories can be summarized as
\begin{subequations}\label{UIRcat}
\begin{align}
    c&=\pm i\rho,~~\rho\in\mathbb{R}^+~~~~\text{principal series}\\
    l=0:~~c&\in\left(-\frac{3}{2},\frac{3}{2}\right)~~~~~~~~~~~~\text{complementary series}\\
    c&=\pm \frac{3}{2}~~~~~~~~~~~~~~~~~~~\text{discrete series (exceptional type I)}
\end{align}
\end{subequations}

In the coordinate system \eqref{metric}, the action of the dilatation is easier to understand at the late-time boundary, which is reached in the limit $\eta\to 0$. For scalar fields the scaling weight has the following dependence on mass
\begin{align}
    \text{for}~~l=0:~~c^2=\frac{d^2}{4}-\frac{m^2}{H^2}.
\end{align}
Leaving the further details to section \ref{sec:Wavefunc review}, for a free real scalar field with mass $m$ on $dS_4$, which satisfies Bunch Davies initial conditions and approaches a given late-time field profile $\Phi_{\vec{k}}$, where $\vec{k}$ refers to momentum in Fourier space, the classical modes  behave as\footnote{$\tilde{H}^{(1)}_\rho(x)\equiv\tilde{J}_\rho(x)+i\tilde{Y}_\rho(x)=e^{-\rho\pi/2}H^{(1)}_{i\rho}(x)$ is one of the solutions to what the Bessel's equation turns into when the parameter becomes purely imaginary while the argument of the function is real \cite{NIST:DLMF}.} 
 \begin{equation}\label{eqn:classical modes}
    \phi_{\vec{k}}=\Phi_{\vec{k}}\frac{v_k(\eta)}{v_k(\eta_0)},~~~v_k(\eta)=\begin{cases}|\eta|^{3/2}\tilde{H}^{(1)}_\rho(k|\eta|),~~\rho^2=\frac{m^2}{H^2}-\frac{9}{4} & \text{if } m>\frac{3}{2}H\\
    |\eta|^{3/2}H^{(1)}_\nu(k|\eta|),~~\nu^2=\frac{9}{4}-\frac{m^2}{H^2} & \text{if } 0<m<\frac{3}{2}H\\
    |\eta|^{3/2}H^{(1)}_\frac{3}{2}(k|\eta|), & \text{if } m=0.
    \end{cases}
 \end{equation}

For the scalar fields on the fixed de Sitter background, the scaling weight $c$ gets identified with the order of the Hankel function in \eqref{eqn:classical modes}. Principal series representations capture heavy matter fields in any dimensions while complementary series capture light. Here light and heavy is determined in terms of the de Sitter scale $H=l^{-1}_{dS}$, where $H$ is the Hubble parameter which has a constant value for de Sitter. Scalar fields with masses $m>\frac{3}{H}$ are heavy and correspond to principal series, while scalar fields of masses $0<m<\frac{3}{H}$ correspond to complementary series. The massless scalar belongs to the exceptional series type-I category, which we will shortly refer to as discrete series from now on.

In what follows we will mostly leave complementary series out of our discussion. Focusing on the principal and discrete series is enough to have an idea about matter fields and gauge fields and exhibits the contrasting features of the two from the perspective of cluster decomposition.

\section{Review of two-point functions on the $S^4$ and $dS_4$}
\label{sec:Review of two-point functions on the $S^4$ and $dS_4$}
In this section we review the techniques of calculating two-point functions on $S^4$ and $dS_4$. The two geometries are related by a Wick rotation. One can Wick rotate global $dS_4$
\begin{align}
   ds^2_{global~dS_4}=-dT^2+\frac{1}{H^2}\cosh^2(HT)\left[d\theta_2^2+\sin^2\theta_2 d\Omega^2_2\right]
\end{align}
to $S^4$ of radius length $H^{-1}$ with the metric
\begin{align}
   \label{eqn:S4metric} ds^2_{S^4}=\frac{1}{H^2}\left[d\theta_1^2+\sin^2\theta_1\left[d\theta_2^2+\sin^2\theta_2d\theta_3^2+\sin^2\theta_2\sin^2\theta_3d\theta_4^2\right]\right]
\end{align}
via
\begin{align}\label{eqn:analytic cont}T\to ir,~~\text{followed by}~~\theta_1=Hr+\frac{\pi}{2}.
\end{align}

\subsection{Two-point functions on $S^4$ and the zero modes}
\label{sec:zeromodes}
Unless properly treated, zero-modes can become problematic. Here to understand the problems that may arise we repeat the calculation of \cite{Folacci_1992,anninos2023discreet} of the two point function for a scalar field on four sphere, this is the Euclidean version of the problem on $dS_4$ in global coordinates. Our main quest in this section is to understand if the zero modes are a problem in the case of principal series as they are for discrete series. 

A free massive scalar field 
\begin{align}
    \label{eqn:action SE}S_E=\frac{1}{2}\int\sqrt{g}d^4\theta \left[\left(\Box_{S^4}\phi\right)^2+m^2\phi^2\right]
\end{align}
on the four sphere of radius length $H^{-1}$ with the metric \eqref{eqn:S4metric} satisfies the following equation of motion
\begin{align}
\label{eqn:sphereeom}\Box_{S^4}\phi=\frac{m^2}{H^2}\phi
\end{align}
where $\Box_{S^4}$ is the d'Alembertian on the four sphere. This looks like an eigenvalue problem however, instead of mass, we will use the scaling dimension, which in four dimensions reads
\begin{align}
    \Delta=\frac{3}{2}+c,~~\text{with}~~c^2=\frac{9}{4}-\frac{m^2}{H^2}.
\end{align}
We will pick $\rho$ to denote the positive root, such that $c=\pm\rho$ and $\rho$ is purely imaginary for principal series fields with masses in the range $\frac{m^2}{H^2}>\frac{9}{4}$. The conversion of mass to scaling dimension works as
\begin{align}
    \label{eqn:m to Delta} \frac{m^2}{H^2}=-\Delta\left(\Delta-3\right).
\end{align}
Equation \eqref{eqn:sphereeom} suggests that we can expand the field in terms of the eigenfunctions of the d'Alembertian operator. Labeling the eigenfunctions of the $S^4$ d'Alembertian by $\lambda_n$ and a label $i$ related with degeneracy, as in \cite{Folacci_1992}, we have
\begin{align}
    \label{eqn:S4 eigenfuncts} \Box_{S^4}\phi^i_n=-\lambda_n \phi^i_n,~~&i=1,\dots,d_n
\end{align}
where 
\begin{align}\label{eqn:S4 eigenvalues}\lambda_n=H^2n(n+3),~~&n=0,1,2,\dots,\\
\nonumber&d_n=\frac{1}{6}(n+1)(n+2)(n+3),\end{align} 
$d_n$ denotes the degeneracy of each value. These eigenfunctions form an orthonormal basis for scalar fields with \cite{Folacci_1992}
\begin{subequations}\label{eqn:S4orthonorm}
\begin{align}
    \int \sqrt{g}d^4\theta\phi^i_n\phi^j_m&=\delta_{ij}\delta_{nm},\\
    \sum_n\sum_{i=1}^{d_n} \phi^i_n(\theta)\phi^i_n(\theta')&=\delta^{(4)}(\theta-\theta').
\end{align}
\end{subequations}
Let us expanding the field $\phi$ in terms of the eigenfunctions of the sphere d'Alembertian,
\begin{align}
    \label{eqn:spherical expansion} \phi=\sum_n a_n\phi^i_n.
\end{align}
Inserting the field expansion \eqref{eqn:spherical expansion} into the action \eqref{eqn:action SE}, integrating by parts and rewriting the mass in terms of the scaling dimension we arrive at
\begin{align}
\label{eqn:SEafterexpansion}    S_E&=-\frac{1}{2}\int \sqrt{g}d^4\theta\left[\phi\Box_{S^4}\phi+\Delta(\Delta-3)H^2\phi^2\right]\\
    &=-\frac{1}{2}\sum_n \left[\Delta(\Delta-3)H^2-\lambda_n\right]a^2_n.
\end{align}
This form of the action simplifies the two-point function calculation by turning the integration over field configuration into a Gaussian integral over $a_n$. 

The two-point function on the sphere can be computed via the following path integral
\begin{align}
G(\theta,\theta')&=\langle\phi(\theta)\phi(\theta')\rangle\\
    &=\frac{1}{\mathcal{N}}\int\mathcal{D}\phi \phi(\theta)\phi(\theta') e^{-S}
\end{align}
where the normalization coefficient is defined as
\begin{align}
    \mathcal{N}\equiv \int\mathcal{D}\phi e^{-S},
\end{align}
and the measure denotes the measure on field space. By expanding the field in terms of $S^4$ eigenfunctions via \eqref{eqn:spherical expansion} we have treaded in the information on field configuration in terms of the coefficients $a_n$. Accordingly, the measure is now given by
\begin{align}
    \mathcal{D}\phi=\prod_n da_n.
\end{align}
Notingthat
\begin{align}
\label{eqnphiphiexp}\phi(\theta)\phi(\theta')=\sum_n\sum_m a_n a_m \phi_n^i(\theta)\phi^j_m(\theta'),
\end{align}
and the fact that Gaussian integrals with odd arguments vanish, the only nonvanishing contribution from this double sum will come for the case of $m=n$. This contribution to the numerator for the two-point function comes in the form
\begin{align}
    \label{eqn:2pnum}
    \int \prod_l da_l\sum_n a_n^2\phi_n^i(\theta)\phi^j_n(\theta')e^{-\frac{1}{2}\left[\Delta\left(\Delta-3\right)H^2-\lambda_l\right]a_l^2}\end{align}
     Expanding the products carefully we have   \begin{align}\label{eqn:2pnum2}\nonumber&\sum_n\phi_n^i(\theta)\phi^j_n(\theta')\left(\int da_n a_n^2 e^{-\frac{1}{1}\left[\Delta\left(\Delta-3\right)H^2-\lambda_n\right]a_n^2}\right)\left(\int\prod_{l\neq n}da_l e^{-\frac{1}{1}\left[\Delta\left(\Delta-3\right)H^2-\lambda_l\right]a_l^2} \right)\\
&=\sum_n\phi_n^i(\theta)\phi^j_n(\theta')\left(\frac{1}{\Delta\left(\Delta-3\right)H^2-\lambda_n}\sqrt{\frac{2\pi}{\Delta\left(\Delta-3\right)H^2-\lambda_n}}\right)\left(\int\prod_{l\neq n}da_l e^{-\frac{1}{1}\left[\Delta\left(\Delta-3\right)H^2-\lambda_l\right]a_l^2}\right).
\end{align}
The factors of $\sqrt{\frac{2\pi}{\Delta\left(\Delta-3\right)H^2-\lambda_n}}$ in the numerator cancel with those in the denominator. Expressing $\lambda_n$ explicitly as well, we reach
\begin{align}
\label{eqn:twopointonS4}G_{S^4}(\theta,\theta')=\sum_{n=0}\frac{1}{H^2}\frac{\phi_n^i(\theta)\phi^j_n(\theta')}{\Delta\left(\Delta-3\right)-n(n+3)}.
\end{align}

Let us explore what happens to this expression for each category of representations.
\subsubsection{The trouble with the zero-mode contribution}
\label{subsubsec:The trouble with the zero-mode contribution}
For bosonic discrete series representations the scaling dimension is a nonnegative integer
\begin{align}
    \Delta=0,1,2,3,\dots.
\end{align}
At a first glance the zero mode, $n=0$, in \eqref{eqn:twopointonS4} is problematic for the cases of $\Delta=0$ and $\Delta=3$, which arise for the case of a massless scalar on $dS_4$. This is one observation that points towards the importance of how to treat the zero mode which has been discussed many times in the literature with specific considerations on the massless scaler on de Sitter, especially in \cite{Higuchi:1991tm} with emphasis on its removal in the case of non-interacting theories. This observation also makes one wonder if other modes in the sum will blow up for other dimensions among the discrete series representations? 

The $S^4$ harmonics we used in the expansion \eqref{eqn:spherical expansion} only capture scalar symmetric traceless spherical harmonics. Thus within our analysis we can only discuss scalar fields. There are symmetric traceless spherical harmonics for other spins as well, both for integer \cite{Higuchi:1986wu} and half integer \cite{Letsios:2020twa} spins. The analysis has to be carried out accordingly for these spins. Shift symmetric scalars on $dS^4$ are the candidate discrete series representations. These are the massless scalar and tachyonic scalars. To be more precise these fall under Exceptional series type-I representations, while Exceptional series type-II on $dS_4$ and $dS_2$ are properly recognized as discrete series representations and they capture gauge fields with non-zero spin.

Among the other examples to scalar discrete series representations, the scalar tachyons \cite{Bros_2010}, have their masses parametrized as 
\begin{align}
    \label{eqn:tachyon mass}m^2_t=-t(t+3)H^2,~~t\in\mathbb{R}^+.
\end{align}
In this case the denominator is
\begin{align}
    \Delta(\Delta-3)-n(n+3)=t(t+3)-n(n+3)
\end{align}
which will blow up for $n=t$ and $n=-3-t$, where the later possibility is excluded from the sum in \eqref{eqn:twopointonS4}.

At a more careful look, the massless scalar action accommodates shift symmetry. The scalar field can be shifted by a constant $\phi\to\phi+\phi_0$ and nothing will change in the action. The equation of motion for a massless scalar,
\begin{equation}
    \Box_{S^4}\phi=0,
\end{equation}
allows for constant solutions however these are not physical solutions since the constant value of the field can always be shifted to some other value. Accordingly all constant solutions for the massless scalar are gauge and should be removed. This can be understood as fixing a particular gauge. The constant solution in terms of the decomposition over the spherical harmonics corresponds to the zero mode. This shift symmetry is present for all the tachyonic scalars and the same argument holds. 

The scalars with other values of mass fall under principal and complementary series where the ranges for the scaling dimension \begin{align}\Delta_{principal}&=\frac{3}{2}\pm i\rho,~~\text{ with}~~\rho\in\mathbb{R}^+,\\ \Delta_{complementary}&=\frac{3}{2}+\nu~~\text{ with}~~\nu\in\left(-\frac{3}{2},\frac{3}{2}\right),\end{align}
do not cause any problems for the denominator in \eqref{eqn:twopointonS4}. However, for these cases the constant solution is never a solution to the equation of motion
\begin{equation}
    \Box_{S^4}\phi=\frac{m^2}{H^2}\phi,
\end{equation}
Hence it still makes sense to remove the zero mode from the sum in \eqref{eqn:twopointonS4}.

Thus the better defined two-point function is 
\begin{align}
\label{eqn:twopointonS4_better}G'_{S^4}(\theta,\theta')=\sum_{n=1}\frac{1}{H^2}\frac{\phi_n^i(\theta)\phi^j_n(\theta')}{\Delta\left(\Delta-3\right)-n(n+3)}.
\end{align}
which is \eqref{eqn:twopointonS4} with the zero-mode removed and we use prime to emphasize this point. In the case of massless and tachyonic scalars, the zero-mode wouldn't have contributed if one handled the gauge fixing properly from the beginning and in the case of scalars of other mass zero-mode doesn't contribute because it is automatically not a solution to the equations of motion.

In what follows we will compute $dS_4$ two-point functions in position space and in field space. In position space two-point functions we will point out that removal of the zero-mode does not change the result on the clustering properties of a particular category of representations. This guarantees that the curious case of the non-locality of the discrete series two-point function really is a feature of the discrete series representation and not a gauge artifact. In field space two-point functions we will point out to further reasons that require the removal of the zero-mode for any representation category.

\subsection{The wavefunction and two-point functions on $dS_4$}
\label{sec:Wavefunc review}

Our main discussion branches into two venues, the concept of cluster decomposition in physical space and in field space. Our main method will be to use the Wavefunction formalism and we will see that in both discussions two-point functions play an important role. So let us first summarize the wavefunction and how to compute the correlation functions from it.

The wavefunction is a functional of a given field profile $\Phi$, specified at a specific time $\eta_0$, 
\begin{align}\label{eqn:wavefunc:def}
     \Psi[\Phi,\eta_0] \equiv \int^{\phi(\eta_0)=\Phi}_{\lim_{\eta\to-\infty(1+i\epsilon)}\phi(\eta)\to 0} \mathcal{D}\phi e^{iS[\phi]}.
\end{align}
For us $\eta_0$ is the late-time. From now on, we will work in conformal planar patch coordinates where the metric is that of \eqref{metric}, and the late-time corresponds to $\eta\to 0$.
The lower limit implies the Bunch-Davies initial condition \cite{Bunch:1978yq} where the field is to behave as if on flat spacetime at early times. In practice a semiclassical approximation via
\begin{align}
    \label{eqn:wfn:appr}
    \Psi[\Phi,\eta_0] \sim e^{iS_{onshell}[\Phi,\eta_0]},
\end{align}
is more manageable then calculation of the formal expression in \eqref{eqn:wavefunc:def}. 

In considering correlation functions, the calculation turns out to be easier to carry on in momentum space. One can go back and forth between position space and momentum space via Fourier transformation. In going over to the momentum space, the Fourier modes are defined by
 \begin{equation}\label{eqn:Psi_k}
     \phi(\eta,\vec{x})=\int\frac{d^dk}{(2\pi)^d}\phi_{\vec{k}}(\eta)e^{i\vec{k}\cdot\vec{x}}.
 \end{equation}
 For a real scalar field, the Fourier modes are required to satisfy $\phi_{\vec{k}}^*=\phi_{-\vec{k}}$ and they behave as given in \eqref{eqn:classical modes} with respect to the Bunch Davies initial and the late-time profile $\Phi_k$ which is the Fourier momentum space version of the position-space late-time field profile $\Phi$. The onshell action for the free scalar field with the boundary conditions in \eqref{eqn:wavefunc:def} is
\begin{align} \label{onshellintro} S_{onshell}[\Phi,\eta_0]=-\frac{1}{2}\int \frac{d^3k}{(2\pi)^{3}}a^2(\eta_0)\frac{v'_k(\eta_0)}{v_k(\eta_0)}|\Phi_{\vec{k}}|^2,\end{align}
 with $v_k$ being the modes we introduced in equation \eqref{eqn:classical modes} and the scale factor for the conformal planar patch metric is $a(\eta_0)=\frac{1}{H|\eta_0|}$.
 
 All these considerations lead to the following form of the wavefunction \cite{Guven:1987bx}
\begin{align}
 \label{eqn:wavefunction format}   \Psi[\Phi,\eta_0]=\mathcal{N}(\eta_0)exp\left[-\frac{1}{2} \int \frac{d^dk}{(2\pi)^d}\mathcal{P}(k, |\eta_0|)\Phi_{\vec{k}}\Phi_{-\vec{k}} \right],
\end{align}
with the normalization being
\begin{equation}
    \frac{1}{|\mathcal{N}(\eta_0)|^2}=\int\mathcal{D}\Phi_{\vec{k}}|\Psi(\Phi_{\vec{k}})|^2.
\end{equation}
 Taking the reality condition on the field and its implication for the Fourier modes into account, our measure is
\begin{equation}
\label{eqn:Psi measure}\mathcal{D}\Phi_{\vec{k}}=\prod_{\vec{k}\in {R}^{+d}}d\Phi_{\vec{k}}.
\end{equation}
 
 Being interested in only field profiles, and not conjugate momenta, what goes into the calculation of field correlations is the amplitude squared of the wavefunction, which in itself can be parametrized in the convention of \cite{Anninos:2011kh} as,
 \begin{equation}\label{eqn:intro beta}
|\Psi(\Phi_{\vec{k}})|^2=|\mathcal{N}_{\vec{k}}(\eta)|^2e^{-2\int\frac{d^dk}{(2\pi)^d}\beta(\vec{k},\eta)\Phi_{\vec{k}}\Phi_{\vec{k}}^*}.
\end{equation}
The two parametrizations are related by
\begin{equation}
    \beta(k,\eta)=\frac{1}{4}\left(\mathcal{P}(k,\eta)+\mathcal{P}^*(k,\eta)\right).
\end{equation}
 
Notice that the wavefunction of a free field on de Sitter is a Gaussian in the form of
\begin{equation}
    |\Psi(\Phi_{\vec{k}})|^2=e^{-\frac{1}{2}\int d^dq\Phi_{\vec{k}}\mathcal{A}(\vec{k},\vec{q})\Phi_{\vec{q}}},
\end{equation} 
we can make use of the identity
\begin{equation}\label{eqn:twopointgaussianintegration}
\langle\mathcal{O}_i\mathcal{O}_j\rangle=\frac{\int d\mathcal{O}_1\dots\int d\mathcal{O}_N e^{-\frac{1}{2}\vec{\mathcal{O}}^T\cdot\mathcal{A}\cdot\vec{\mathcal{O}}}\mathcal{O}_i\mathcal{O}_j}{\int d\mathcal{O}_1\dots\int d\mathcal{O}_N e^{-\frac{1}{2}\vec{\mathcal{O}}^T\cdot\mathcal{A}\cdot\vec{\mathcal{O}}}}=\mathcal{A}^{-1}(i,j).
\end{equation}
As computed in \cite{Sengor:twopoint}, for our case
$\mathcal{A}(\vec{k},\vec{q})$ and its inverse are
\begin{subequations}
    \begin{align}
        \mathcal{A}(\vec{k},\vec{q})&=\frac4{(2\pi)^d}\beta(\vec{k},\eta)\delta^{(d)}(\vec{q}+\vec{k}),\\
        \mathcal{A}^{-1}(\vec{k},\vec{q})&=\frac{(2\pi)^d}{4\beta(\vec{k},\eta)}\delta^{(d)}(\vec{q}+\vec{k}).
    \end{align}
\end{subequations}  
Accordingly, the late-time two point functions for the field in momentum space have the following functional form
\begin{subequations}
    \label{twopoint_form}
\begin{align}
\langle\Phi_{\vec{k}}\Phi_{\vec{k}'}\rangle&=\frac{(2\pi)^d}{2\left(\mathcal{P}(k,\eta)+\mathcal{P}^*(k,\eta)\right)}\delta^{(d)}\left(\vec{k}+\vec{k}'\right)\\
&=\frac{(2\pi)^d}{8\beta(\vec{k},\eta)}\delta^{(d)}(\vec{k}+\vec{k}').
\end{align}
\end{subequations}
Below we list the explicit expressions of the functions $\mathcal{P}$ and $\beta$ for each category of representations which we will refer to in later sections.

At late times $\mathcal{P}$ is a function of momentum, and carries information about the scaling behavior of correlation functions. The exact functional dependence of $\mathcal{P}$ depends on the representation category under consideration.  As far as scalar fields are concerned the expressions in the late-time limit are as follows \cite{Isono:2020qew,Sengor:twopoint},
 \begin{subequations}\label{eqn:thePs}
     \begin{align}
     \label{Pprinc}
\mathcal{P}^{({\rm Prin})}(k)&=-\frac{|\eta_0|^{-d}}{H^{d-1}}\frac{\left(i\frac{d}{2}+\rho\right)-\left(i\frac{d}{2}-\rho\right)e^{-\rho\pi-2i\gamma_\rho}\left(\frac{k''|\eta_0|}{2}\right)^{2i\rho}}{1-e^{-\rho\pi-2i\gamma_\rho}\left(\frac{k''|\eta_0|}{2}\right)^{2i\rho}},\\
 \label{Pcomp} \mathcal{P}^{{(\rm Comp)}}(k)&=\frac{1}{|\eta_0|^dH^{d-1}}\frac{\frac{i\left(1+i \cot(\nu\pi)\right)}{\Gamma(\nu+1)}\left(\frac{d}{2}+\nu\right)\left(\frac{k|\eta_0|}{2}\right)^{2\nu}+\frac{\Gamma(\nu)}{\pi}\left(\frac{d}{2}-\nu\right)}{\frac{i\Gamma(\nu)}{\pi}-\frac{1+i\cot(\nu\pi)}{\Gamma(\nu+1)}\left(\frac{k|\eta_0|}{2}\right)^{2\nu}}\\      \mathcal{P}^{({massless})}_{d=3}(k)&=\frac{3\pi}{|\eta_0|^3H^2}\frac{1}{\Gamma(\frac{3}{2})\Gamma(\frac{3}{2}+1)\left(\frac{k|\eta_0}{2}\right)^{-3}+i\pi},
     \end{align}
 \end{subequations}
 where for the case of discrete series, the allowed dimensions are fixed to be $d=\{1,3\}$ and here we are focusing on $d=3$.

At the late-time, for principal series and for massless scalar which represents the discrete series representations in the scalar sector, $\beta$ has the following behavior respectively
\begin{subequations}
\label{eqn:beta categoric beh}
    \begin{align}
       &\beta^{princ}(k,\eta_0)=\frac{1}{4}\frac{2^{2i\rho+1}e^{2i\gamma_\rho}(e^{2\pi\rho}-1)H^{1-d}\rho}{(2^{2i\rho}e^{2i\gamma_\rho+\pi\rho}-(k|\eta_0|)^{2i\rho})(2^{2i\rho}e^{2i\gamma_\rho}-e^{\pi\rho}(k|\eta_0|)^{2i\rho})}\frac{k^{2i\rho}}{|\eta_0|^{d-2i\rho}}\\
        \nonumber&\beta^{comp}(k,\eta_0)=[4^{-\nu} H^{1-d}\pi\Gamma^2(1+\nu)]\frac{k^{2\nu}}{|\eta_0|^{d-2\nu}}\\
         &\times        \left[\Gamma(\nu)\Gamma(1+\nu)\left[\Gamma(\nu)\Gamma(1+\nu)-\frac{2\pi\cot{(\pi\nu)}}{4^{\nu}}(k|\eta_0|)^{2\nu}\right]+\frac{\pi^2\csc^2{(\pi\nu)}}{4^{2\nu}}(k|\eta_0|)^{4\nu}\right]^{-1}\\
         &\beta^{massless}_{d=3}(k,\eta_0)=\frac{1}{2H^2}\frac{1}{\frac{1}{9}\left(k|\eta_0|\right)^6+1}k^3
    \end{align}
\end{subequations}
 For superhorizon modes $k|\eta|\ll 1$, $\beta$ shows the same behaviour for any mass in the late-time limit   \cite{Roberts:2012jw}
\begin{equation}
    \beta_{k|\eta|\ll1}\sim\frac{k^{d-2\Delta_-}}{|\eta_0|^{2\Delta_-}}
\end{equation}
upto a proportionality constant that depends on the mass and dimensions.

\section{ Cluster decomposition in position space}
\label{sec:clust dec in position space}

For quantum fields on curved space one of the signatures of locality is the two-point function exhibiting a power law decay at large distances in the late-time limit \cite{Benincasa:2022gtd}. In this section we will review the discussion following wavefunction methods initially introduced in \cite{Hartle:1983ai} and using recent results of \cite{Sengor:twopoint,Sengor:particles} where the late-time two-point functions are organized with respect to representation theoretic categories.

With the formalism reviewed in section \ref{sec:Wavefunc review},  the late-time two-point functions for scalar fields in each category come out as follows \cite{Sengor:twopoint} 
\begin{subequations}
    \begin{align}
    \label{2pointPrinc}
\langle\Phi^{Princ}_{\vec{k}}\Phi^{Princ}_{\vec{k}'}\rangle
=&
 \frac{ (2\pi|\eta_0|)^dH^{d-1}}{4\rho \sinh(\rho\pi)} \delta^{(d)}(\vec{k}+\vec{k}')\times
\nonumber
\\
& \ \ \ \   \times
\left[2\cosh(\rho\pi)-e^{2i\gamma_\rho}\left(\frac{k|\eta_0|}{2}\right)^{-2i\rho}-e^{-2i\gamma_\rho}\left(\frac{k|\eta_0|}{2}\right)^{2i\rho}\right],
\\
\nonumber\langle\Phi^{Comp}_{\vec{k}}\Phi^{Comp}_{\vec{k}'}\rangle&=\frac{\pi}{4}(2\pi|\eta_0|)^dH^{d-1}\delta^{(d)}(\vec{k}+\vec{k}')\times\\
\label{2pointComp}&\ \ \ \ \times\left[\frac{1+\cot^2(\nu\pi)}{\Gamma^2(\nu+1)}\left(\frac{k|\eta_0|}{2}\right)^{2\nu}+\frac{\Gamma^2(\nu)}{\pi^2}\left(\frac{k|\eta_0|}{2}\right)^{-2\nu}-\frac{2\cot(\nu\pi)}{\nu\pi}\right],\\
\label{2pointDisc}\langle\Phi^{Discr}_{\vec{k}}\Phi^{Discr}_{\vec{k}'}\rangle&=
\frac{ \pi (2\pi)^3 }{4}H^{2}|\eta_0|^3\delta^{(3)}(\vec{k}+\vec{k}')\left[
\frac{\Gamma^2(\frac{3}{2})}{\pi^2} \left(\frac{ k|\eta_0|}{2} \right)^{-3}
+\frac{1}{\Gamma^2(\frac{5}{2})} \left(\frac{ k|\eta_0|}{2} \right)^{3}\right].
    \end{align}
\end{subequations}
Note that the constant term proportional to $\cot(\nu\pi)$ that is present in the case of complementary series, drops out for the discrete series case, which here implies the massless scalar at $d=3$.

Going back to position space involves integrals of the form
\begin{equation}
    \langle \Phi(\vec{x})\Phi(\vec{y})\rangle=\int \frac{d^dk}{(2\pi)^d}\frac{d^dk'}{(2\pi)^d}\langle\Phi_{\vec{k}}\Phi_{\vec{k}'}\rangle e^{i(\vec{k}\cdot\vec{x}+\vec{k}'\cdot\vec{y})}.
\end{equation}
Considering the form of the field two-point functions as given in \eqref{twopoint_form}, these integrals simplify to
\begin{equation}
    \langle \Phi(\vec{x})\Phi(\vec{y})\rangle=\int \frac{d^dk}{(2\pi)^d}\frac{e^{i\vec{k}\cdot(\vec{x}-\vec{y})}}{2\left(\mathcal{P}+\mathcal{P}^*\right)}=\int \frac{d^dk}{(2\pi)^d}\frac{e^{i\vec{k}\cdot(\vec{x}-\vec{y})}}{8\beta(k,|\eta_0|)}.
\end{equation}
For $d=3$, we are faced with integrals of the form
\begin{equation}\label{eqn:d3xspace}
    \langle \Phi(\vec{x})\Phi(\vec{y})\rangle=\frac{1}{4}\int^\infty_0 \frac{dk}{(2\pi)^2}\frac{k\sin(k|\vec{x}-\vec{y}|)}{|\vec{x}-\vec{y}|\beta(k,|\eta_0|)}.
\end{equation}
Performing these integrals are not easy, but we are mainly interested in the $|\vec{x}-\vec{y}|$ dependence and if whether the $k$- integration converges or not. For some intuition with the $|\vec{x}-\vec{y}|\to \infty$ limit, let's consider a redefinition of variables as
\begin{equation}
    \label{eqn:sec2 change of var} w\equiv k|\vec{x}-\vec{y}|.
\end{equation}
This change of variables leads us to
\begin{equation}
    \langle \Phi(\vec{x})\Phi(\vec{y})\rangle=\frac{1}{4(2\pi)^2}\frac{1}{|\vec{x}-\vec{y}|^3}\int^\infty_0 dw\frac{w\sin(w)}{\beta(\frac{w}{|\vec{x}-\vec{y}|},|\eta_0|)}.
\end{equation}
For this calculation it is convenient to express $\beta$ as follows
\begin{subequations}
    \begin{align}
        \label{eqn:beta2ptPrinc}\frac{1}{\beta_{Princ}}&=\frac{ 2(|\eta_0|)^3H^{2}}{\rho \sinh(\rho\pi)}\times \\       
\nonumber\times&\left[2\cosh(\rho\pi)-e^{2i\gamma_\rho}|\vec{x}-\vec{y}|^{2i\rho}\left(\frac{w|\eta_0|}{2}\right)^{-2i\rho}-e^{-2i\gamma_\rho}|\vec{x}-\vec{y}|^{-2i\rho}\left(\frac{w|\eta_0|}{2}\right)^{2i\rho}\right],\\
\frac{1}{\beta_{Comp}}&=2\pi (|\eta_0|)^3H^2\times\\
\nonumber&\times\left[\frac{1+\cot^2(\nu\pi)}{\Gamma^2(\nu+1)}|\vec{x}-\vec{y}|^{-2\nu}\left(\frac{w|\eta_0|}{2}\right)^{2\nu}+\frac{\Gamma^2(\nu)}{\pi^2}|\vec{x}-\vec{y}|^{2\nu}\left(\frac{w|\eta_0|}{2}\right)^{-2\nu}-\frac{2\cot(\nu\pi)}{\nu\pi}\right],\\
\frac{1}{\beta_{Discr}}&= H^3|\eta_0|^3\left[\frac{1}{2}|\vec{x}-\vec{y}|^3\left(\frac{w|\eta_0|}{2}\right)^{-3}+\frac{32}{9}|\vec{x}-\vec{y}|^{-3}\left(\frac{w|\eta_0|}{2}\right)^{3}\right]    \end{align}
\end{subequations}
Suppressing the constants as $c^{category}_i$, the expressions for each category become 
\begin{subequations}\label{2pt_position space w}\begin{align}
 \langle \Phi^{P}(\vec{x})\Phi^{P}&(\vec{y})\rangle=c^P_1\frac{|\eta_0|^{3}}{|\vec{x}-\vec{y}|^{3}}\int^\infty_0 dw w\sin(w)\\
\nonumber &+c^P_2\frac{|\eta_0|^{3-2i\rho}}{|\vec{x}-\vec{y}|^{3-2i\rho}}\int^\infty_0 dw w^{1-2i\rho}\sin(w)+c^P_3\frac{|\eta_0|^{3+2i\rho}}{|\vec{x}-\vec{y}|^{3+2i\rho}}\int^\infty_0 dw w^{1+2i\rho}\sin(w),\\
 \nonumber \langle \Phi^{C}(\vec{x})\Phi^{C}&(\vec{y})\rangle=c^C_1 \frac{|\eta_0|^{3}}{|\vec{x}-\vec{y}|^{3}}\int^\infty_0 dw w\sin(w)\\
 &+c^C_2\frac{ |\eta_0|^{3-2\nu}}{|\vec{x}-\vec{y}|^{3-2\nu}}\int^\infty_0 dw w^{1-2\nu}\sin(w)+c^C_3\frac{ |\eta_0|^{3+2\nu}}{|\vec{x}-\vec{y}|^{3+2\nu}}\int^\infty_0 dw w^{1+2\nu}\sin(w),\\
\langle \Phi^{D}(\vec{x})\Phi^{D}&(\vec{y})\rangle=c_1^D\int^\infty_0 dw w^{-2}\sin(w)+c^D_2\frac{|\eta_0|^6}{|\vec{x}-\vec{y}|^{6}}\int^\infty_0dw w^4\sin(w). 
\end{align}\end{subequations}

Cluster decomposition is a statement on the behavior of two-point functions at large distances. Accordingly we are interested in the behaviour of \eqref{2pt_position space w} in the limit $|\vec{x}-\vec{y}|\to\infty$. At first sight, both principal and complementary series exhibit power law decay and satisfies cluster decomposition. Principal series two-point functions satisfy cluster decomposition because the real part decays as $|\vec{x}-\vec{y}|^{-3}$. For the complementary series, in our notation $\nu$ is always positive and bounded to be in the range $0<\nu< \frac{3}{2}$ by the unitarity properties of complementary series, which guarantees the power law decay and cluster decomposition. In the case of discrete series, there is a constant piece that will remain in the $|\vec{x}-\vec{y}|\to\infty$  limit, making the discrete series disobey cluster decomposition. 

At a closer look, the terms in \eqref{2pt_position space w} all involve integrals of the form
\begin{equation}\label{eqn:general integr}
    \int^\infty_0 dw w^{1+2n}\sin(w)=-\Gamma\left(2+2n\right)\sin\left(n\pi\right)~~\text{provided}~~-\frac{3}{2}<Re[n]<-\frac{1}{2}.
\end{equation}
Yet every category in \eqref{2pt_position space w} involves at least one integral that is outside the range of \eqref{eqn:general integr} and requires a more careful study. Let's consider these integrals for the principal and discrete series cases in separate sections to justify our claim.

\subsection{Principal series two-point function in position space}
\label{subsec:Principal series two-point function in position space}
For $d=3$, substituting everything into \eqref{eqn:d3xspace},  the scalar principal series two-point function involves the following three integrals
\begin{align}\label{eqn:principal 2pt fourier}
\nonumber\langle\Phi(\vec{x})\Phi(\vec{y})\rangle&=\frac{2\pi H^2 coth(\rho\pi)}{\rho}\frac{|\eta_0|^3}{r}\int^\infty_0 dk k\sin(kr)\\
\nonumber &-\frac{\pi e^{-2i\gamma_\rho}H^2}{\rho\sinh(\rho\pi)}\frac{|\eta_0|^{3-2i\rho}}{2^{-2i\rho}r}\int^\infty_0dk k^{1-2i\rho}\sin(kr)]\\
&-\frac{\pi e^{-2i\gamma_\rho}H^2}{\rho\sinh(\rho\pi)}\frac{|\eta_0|^{3+2i\rho}}{2^{2i\rho}r}\int^\infty_0dk k^{1+2i\rho}\sin(kr)].
\end{align}
These integrals all give $r$-dependent functions at the end of the day. Let's start with the first integral. A purely trigonometric function such as, 
\begin{align}
    \sigma(r,\epsilon)\equiv\frac{1}{2}\left[\int^\infty_0 dk e^{(i-\epsilon)kr}+\int^\infty_0 dk e^{-(i+\epsilon)kr}\right]
\end{align}
where $\epsilon>0$ helps regulate the upper boundary, is easier to integrate. This integral gives
\begin{equation}\label{eqn:sigma r ep}
    \sigma(r,\epsilon)=-\frac{\epsilon}{(i-\epsilon)(i+\epsilon)}\frac{1}{r}.
\end{equation}
The first integral in \eqref{eqn:principal 2pt fourier} can be rewritten as
\begin{align}
    \int^\infty_0 dk k\sin(kr)=-\frac{d}{dr}\sigma(r,0).
\end{align}
By \eqref{eqn:sigma r ep} we have that $\sigma(r,0)=0$, and hence the first term vanishes. 

The other two integrals are both of the form
\begin{equation}
    J(r,\nu)\equiv \int^\infty_0 dk k^{1+i\nu}\sin(kr).
\end{equation}

Considering the integral form of the Gamma function
\begin{align}
    \Gamma(z)\equiv \int^\infty_0t^{z-1}e^{-t}dt,~~Re[z]>0,
\end{align}
using the exponential expansion of the sine function and performing a change of variables one can rewrite $J(r)$ in terms of the Gamma functions as
\begin{align}
    J(r,\nu)&=\frac{\Gamma(2+i\nu)}{2i}\left[\left(\frac{i}{r}\right)^{2+i\nu}-\left(-\frac{i}{r}\right)^{2+i\nu}\right],\end{align}
    which can further be put in a more compact form similar to that of \eqref{eqn:general integr} by expressing factors of $i$ in terms of exponential and using the identities for hyperbolic functions as 
    \begin{align}
   J(r,\nu) &=-\frac{\Gamma(2+i\nu)}{r^{2+i\nu}}i\sinh(\pi\nu/2).
\end{align}
Then one has
\begin{align}
 \int^\infty_0dk k^{1-2i\rho}\sin(kr)&=J(r,-2\rho)=i\frac{\Gamma(2-2i\rho)}{r^{2-2i\rho}}\sinh(\pi\rho),\\ 
 \int^\infty_0dk k^{1+2i\rho}\sin(kr)&=J(r,+2\rho)=-i\frac{\Gamma(2+2i\rho)}{r^{2+2i\rho}}\sinh(\pi\rho).
\end{align}
Everything put together \eqref{eqn:principal 2pt fourier} becomes
\begin{align}
\nonumber\langle\Phi(\vec{x})\Phi(\vec{y})\rangle&=-i\frac{\pi e^{-2i\gamma_\rho}H^2}{2^{-2i\rho}\rho}\frac{|\eta_0|^{3-2i\rho}}{r^{3-2i\rho}}\Gamma\left(2-2i\rho\right)\\
&+i\frac{\pi e^{-2i\gamma_\rho}H^2}{2^{2i\rho}\rho}\frac{|\eta_0|^{3+2i\rho}}{r^{3+2i\rho}}\Gamma\left(2+2i\rho\right)
,
\end{align}
where none of the terms in  exhibit growth at large separation. The principal series late-time two-point function has some oscillatory contributions yet on the whole it decays via power law as $|\vec{x}-\vec{y}|\to\infty$, and hence satisfies cluster decomposition.

\subsection{Discrete series two-point function in position space}
\label{subsec:Discrete series two-point function in position space}
For $d=3$, substituting everything into \eqref{eqn:d3xspace},  the scalar discrete series two-point function involves the following two integrals
\begin{align}
\nonumber\langle\Phi(\vec{x})\Phi(\vec{y})\rangle&=\frac{\pi}{2}\frac{H^2}{|\vec{x}-\vec{y}|}\Bigg[\frac{8\Gamma(3/2)^2}{\pi^2}\int^\infty_0\frac{dk}{(2\pi)^2}\frac{\sin(k|\vec{x}-\vec{y}|)}{k^2}\\
\label{eqn:disctox}&+\frac{|\eta_0|^6}{8\Gamma(5/2)^2}\int^\infty_0\frac{dk}{(2\pi)^2}k^4\sin(k|\vec{x}-\vec{y}|)\Bigg].
\end{align}
Both of these integrals lie outside of the range in \eqref{eqn:general integr} and need some renormalization. Since the second term involves a positive exponent of time, $|\eta_0|^6$, it is subleading next to the first term which comes with $|\eta_0|^0$. So we will focus on the first term.

The difficulty with the first integral is the presence of inverse powers of $k$, which lead to simple poles, and the sine in the numerator, which eventually needs to be evaluated at the upper boundary at infinity. 

Observe that the result of the $k$-integration will be an $r\equiv |\vec{x}-\vec{y}|$ dependent function and define
\begin{equation}
    I(r)=\int^\infty_0 dk \frac{\sin(kr)}{k^2}.
\end{equation}
One clever trick is to differentiate as many times in $r$ as it makes the integration easier. That is if we differentiate twice, we eliminate the $\frac{1}{k^2}$ in the integrand. So let's define 
\begin{equation}
    K(r)\equiv -\frac{d^2}{dr^2}I(r)
\end{equation}
and try to evaluate $K(r)$ first. At this point we just have to worry about how to handle the upper boundary. We expand the sine in terms of exponential function and introduce a factor of $e^{-\epsilon kr}$ with $\epsilon>0$
\begin{equation}
    \kappa(r,\epsilon)\equiv \frac{1}{2i}\left[\int^\infty_0 dk e^{(i-\epsilon)kr}-\int^\infty_0 dk e^{-(i+\epsilon)kr} \right].
\end{equation}
Now at the upper boundary $e^{-\epsilon kr}$ leads to an exponential decay and we have
\begin{equation}
    \kappa(r,\epsilon)=-\frac{1}{2i}\left[\frac{1}{(i-\epsilon)r}+\frac{1}{(i+\epsilon)r}\right]
\end{equation}
and
\begin{equation}
    K(r)=\kappa(r,0)=\frac{1}{r}.
\end{equation}
We can obtain our dersired integral $I(r)$ by integrating $K(r)$ with respect to $r$ twice, which gives
\begin{equation}
    I(r)=-r\ln{r}+d_0 r+d_1
\end{equation}
up to two indeterminate integration constants $d_1$ and $d_2$.

Another clever trick for the first integral is to calculate countour integrals of the form 
\begin{align}
\label{eqn:Cn Sn}    C_n(r,\tau)=\int^\infty_0\frac{\sin kr\cos k\tau}{k^n}dk,~~S_n(r,\tau)=\int^\infty_0\frac{\sin kr\sin k\tau}{k^n}dk
\end{align}
with $r\equiv|\vec{x}-\vec{y}|$, $n=0,1,2$. Then $C_2(r,0)$ will give us the first integral in  \eqref{eqn:disctox}. These integrals are explicitly calculated in \cite{Allen:1985ux} and references within, for similar purposes. The calculation relies on noticing that\cite{Allen:1985ux}
\begin{subequations}
    \begin{align}
        \partial_\tau S_n(r
    ,\tau)&=C_{n-1}(r,\tau),\\
    -\partial_\tau C_n&=S_{n-1}(r,\tau).
    \end{align}
\end{subequations}
The result of interest to us is \cite{Allen:1985ux}
\begin{align}
    C_2(r,0)&=-\frac{1}{2}r \ln[r^2]+c_1+c_2 r
\end{align}
where $c_1$ and $c_2$ are undetermined constants. In agreement with the first method.

Thus the leading behavior of the two-point function in position space at late-times is
\begin{align}
\langle\Phi(\vec{x})\Phi(\vec{y})\rangle&=H^2\Bigg[- \ln[|\vec{x}-\vec{y}|]+\frac{c_1}{|\vec{x}-\vec{y}|}+c_2\Bigg]
\end{align}
and its leading position dependence at late-times involves a Log term. Taking the overall minus sign into account this term is $ln\frac{1}{r}$ which diverges as $r\to\infty$. 

The Euclidean calculation on the sphere taught us that for discrete series scalars one has to be careful about the gauge fixing. We would like to be sure that the logarithmic divergence is a physical property of the discrete series late-time two-point function and not just an artifact of a mishandling of the gauge fixing. As a check we consider the removal of the zero mode in performing the Fourier transform from the momentum space to the position space two-point function in the next section. 

\subsubsection{Removing the zero-mode}
\label{subsubsec:Removing the zero-mode}
In section \ref{subsubsec:The trouble with the zero-mode contribution}, the analytic continuation to the sphere showed us that the zero mode is problematic for the discrete series case and it signals a mishandling of the gauge fixing. In this section we would like to investigate what happens if we remove it by hand in the fourier transform of the momentum-space discrete series two-point function. In practice this means changing the lower integration limit to some small but finite $k_0$. This implies considering the integral
\begin{equation} M(r)=\int^\infty_{k_0} dk \frac{\sin(kr)}{k^2},\end{equation}
instead of $I(r)$ above.

Employing the same procedure to handle the upper limit we define 
\begin{equation}
    N(r)\equiv -\frac{d^2}{dr^2}M(r)
\end{equation}
and
\begin{align}
n(r,\epsilon)&\equiv \frac{1}{2i}\left[\int^\infty_{k_0} dk e^{(i-\epsilon)kr}-\int^\infty_{k_0} dk e^{-(i+\epsilon)kr} \right]\\
    &=-\frac{1}{2i}\left[\frac{e^{(i-\epsilon)k_0r}}{(i-\epsilon)r}+\frac{e^{-(i+\epsilon)k_0r}}{(i+\epsilon)r}\right]
\end{align}
such that 
\begin{equation}
    N(r)=n(r,0)=\frac{\cos(k_0r)}{r}.
\end{equation}
Now we have to integrate $N(r)$ twice with respect to r which leads to
\begin{equation}
    M(r)=-r Ci(k_0 r)+\frac{\sin(k_0r)}{r}+C_0 r+C_1
\end{equation}
where $C_0$, $C_1$ are integration constants and $Ci(k_0 r)$ is the cosine integral which is defined as
\begin{equation}
    Ci(k_0r)=\int^{k_0r}_0\frac{\cos(z)-1}{z}dz+ln(k_0r)+\gamma
\end{equation}
where $\gamma$ is the Euler-Mascheroni constant. Due to the $ln(k_0r)$ term in the cosine integral, the discrete series two-point function $\langle\Phi(\vec{x})\Phi(\vec{y})\rangle$ also has the logarithmic term even after the removal of the zero mode.

To summarize, we find that the late-time leading contribution to the discrete series two-point function for a scalar field with or without the zero modes exhibit logarithmic divergence at large scales where $|\vec{x}-\vec{y}|\to\infty$ and hence violates cluster decomposition. This behavior appears as follows
\begin{align}
\langle\Phi(\vec{x})\Phi(\vec{y})\rangle&=H^2\Bigg[- \ln[|\vec{x}-\vec{y}|]+\frac{c_1}{|\vec{x}-\vec{y}|}+c_2\Bigg]\\
\langle\Phi(\vec{x})\Phi(\vec{y})\rangle'&=H^2\Bigg[- \ln[k_0|\vec{x}-\vec{y}|]+\frac{C_1}{|\vec{x}-\vec{y}|}+C_0-\gamma+\frac{\sin(k_0|\vec{x}-\vec{y}|)}{|\vec{x}-\vec{y}|^2}\\
\nonumber &~~~~~~~~~~-\int^{k_0|\vec{x}-\vec{y}|}_0\frac{\cos(z)-1}{z}dz \Bigg]
\end{align}
where prime indicates that zero modes have been removed by introducing the low momentum cutoff scale $k_0$. Hence this logarithmic divergence is not due to the presence of the zero-modes and it is a physical property of the discrete series. 

\section{Cluster decomposition in field space}
\label{sec:cluster decomposition in field space}

In \cite{Anninos:2011kh}, the analysis on clustering properties of de Sitter has been extended with a discussion on the ultrametricity of the wavefunction which may signal towards the existance of memory on de Sitter. The analysis on ultrametricity relies on the probability distribution of distance in field space. Originally, the definition of distance between two possible configurations in field space and the utrametricity feature of the wavefunction is pointed out in \cite{Anninos:2011kh}  only by considering the case of a massless scalar on $dS_4$. These results are extended to the case of general mass in \cite{Roberts:2012jw}. All together in the discussion, two subtleties arise. These subtleties are related to the handling of the \emph{zero modes} and the \emph{definition of distance in field space}. Here we will revisit the handling of the zero-modes explicitly. As we will review, the well defined distance in field space requires the removal of the zero-mode for all mass. This is in agreement with our analysis of section \ref{sec:zeromodes}. In section \ref{sec:zeromodes} we justified the removal of the zero-mode  for principal series because constant field configuration is not a solution to the equations of motion and in the case of discrete series the constant field configuration corresponds to a gauge degree of freedom. 

\begin{figure}
    \centering
\includegraphics[width=1.2\textwidth]{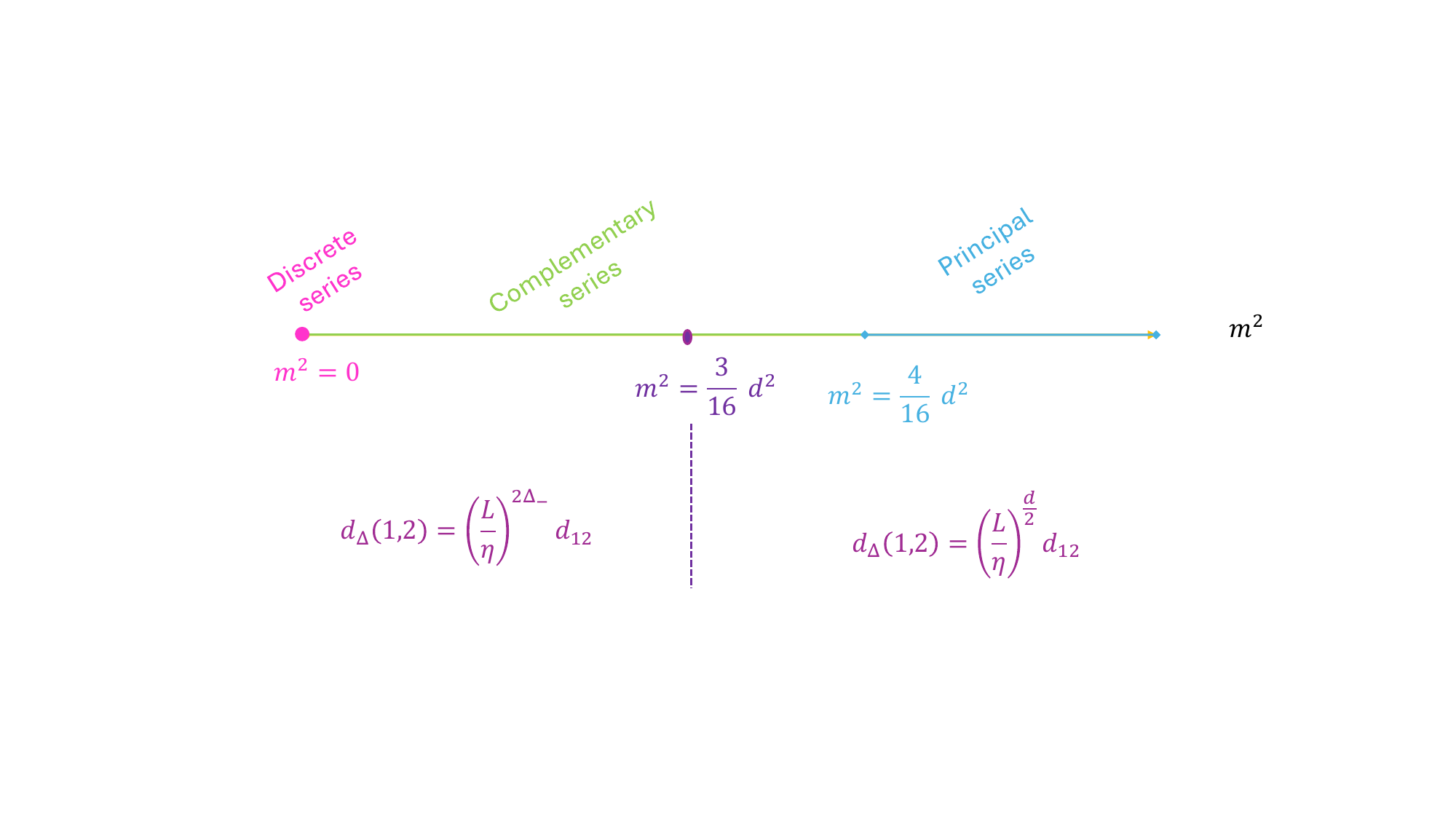}
    \caption{Redefined distance in field space according to \cite{Roberts:2012jw}.}
    \label{fig:roberts figure}
\end{figure}

The revised definition of \cite{Roberts:2012jw} for the distance in field field  when nonzero mass is considered, proposes an adjustment with respect to mass ranges. The main concern of \cite{Roberts:2012jw} in their proposal is on the width of the distribution. The proposed adjustment, summarized in figure \ref{fig:roberts figure}, completely disregards representation theoretic categorization. While the treatment of the zero-modes goes in agreement with what one would expect from the representation theory perspective, the lack of an explicit distinction with respect to representation categories in the revised definition of the distance in field space is the confusing point that requires more attention.  As a first step towards a better understanding we discuss the width of the distribution in section \ref{subsubsec:probability distribution function PD}.

\subsection{The problem with the zero-mode}
\label{subsec:field space problem with the zero-mode}

The bare definition of distance in field space, originally given in \cite{Anninos:2011kh} is introduced by focusing on the massless scalar on $dS_4$ alone. This is the case where zero-modes are expected to be problematic from the start. The analysis we repeated in section \ref{sec:zeromodes} is one clue in this direction. Hence the original definition of \cite{Anninos:2011kh} involves first preparing a redefinition of the field configuration, $\hat{\phi}(\vec{x})$, that is UV finite and has the problematic zero mode removed, via
\begin{align}
    \label{eqn:phihat def} \hat{\phi}(\vec{x})=\int d^dy w(\vec{y})\phi(\vec{x}+\vec{y})-\frac{1}{L^d}\int d^dx \phi(\vec{x}).
\end{align}
Here $\omega(\vec{y})$ is a window function and periodic boundary conditions $x_i \sim x_i+L$ are imposed. Then the bare distance between two field values is defined as
\begin{align}
    \label{eqn:distance def} d_{12}=d[\phi^{(1)},\phi^{(2)}]=\frac{1}{L^d}\int d^dx\left( \hat{\phi}^{(1)}(\vec{x})-\hat{\phi}^{(2)}(\vec{x})\right)^2.
\end{align}
Note that $d_{12}$ only depends on the field profiles and not on position. This is due to the integration over all space. From this perspective $d_{12}$ respects diffeomorphism invariance and is a candidate for a physical observable.

If we express the field profiles in momentum space, 
\begin{align}
    \hat{\phi}^{(i)}(\vec{x})=\int \frac{d^dk}{(2\pi)^d}\hat{\phi}^{(i)}_{\vec{k}}e^{i\vec{k}\cdot\vec{x}},~~~i=1,2,
\end{align}
we will recognize that the two-point function calculations from the earlier sections contribute to $d_{12}$. From here on we drop the hats keeping in mind that the configuration is UV finite but the contribution of the zero-mode is to still to be discussed. The  terms that contribute to $d_{12}$ in momentum space, take on forms such as
\begin{subequations}
    \begin{align}
       \int d^dx \left(\Phi^{(1)}(\vec{x})\right)^2&=\int \frac{d^dk_1}{(2\pi)^d} \left(\Phi^{(1)}_{\vec{k}_1}\Phi^{(1)}_{-\vec{k}_1}\right),\\
       \int d^dx \Phi^{(1)}(\vec{x})\Phi^{(2)}(\vec{x})&=\int \frac{d^dk_1}{(2\pi)^d} \left(\Phi^{(1)}_{\vec{k}_1}\Phi^{(2)}_{-\vec{k}_1}\right),
    \end{align}
\end{subequations}
and overall, $d_{12}$ can be written as follows,
\begin{subequations}
\label{eqn:d12 exps in kspace}
\begin{align}
 \label{eqn:d12_kspace}    d_{12}&=\frac{1}{L^d}\int\frac{d^dk}{(2\pi)^d}\left[\Phi^{(1)}_{\vec{k}}{\Phi}^{(1)}_{-\vec{k}}-{\Phi}^{(1)}_{\vec{k}}{\Phi}^{(2)}_{-\vec{k}}-{\Phi}^{(2)}_{-\vec{k}}{\Phi}^{(1)}_{\vec{k}}+{\Phi}^{(2)}_{\vec{k}}{\Phi}^{(2)}_{-\vec{k}}\right]\\
 \label{eqn:d_kspace}&=\frac{1}{L^d}\int\frac{d^dk}{(2\pi)^d}|\Phi^{(1)}_{\vec{k}}-\Phi^{(2)}_{\vec{k}}|^2,
\end{align}
\end{subequations}
In passing from the first to second line in \eqref{eqn:d12 exps in kspace} we have taken into account that
\begin{align}|\Phi^{(1)}_{\vec{k}}-\Phi^{(2)}_{\vec{k}}|^2=|\Phi^{(1)}_{\vec{k}}|^2-\Phi^{(1)}_{\vec{k}}\Phi^{(2)}_{-\vec{k}}-\Phi^{(2)}_{\vec{k}}\Phi^{(1)}_{-\vec{k}}+|\Phi^{(2)}_{\vec{k}}|^2.\end{align}
We will mainly use the momentum space expressions \eqref{eqn:d12 exps in kspace}. In momentum space the zero-mode contributes purely via $k=0$. Effectively, removing the zero-mode amounts to removing $k=0$ from the limits of integration.

In physical space, that is in position or momentum space,  the problematic contribution of the zero mode to the two-point function is more explicit in the case of the discrete series representation, as we saw in section \ref{sec:zeromodes} via analytical continuation to the sphere. However, in field space the zero mode turns out to be just as explicitly problematic for all representation categories. To emphasize this difference between field space and physical space, let us first calculate the expectation value of distance in field space without the zero-mode removed for each representation category.

In computing $\langle d_{12}(\phi^{(1)},\phi^{(2)})\rangle$, the first thing to notice is that $d_{12}(\phi^{(1)},\phi^{(2)})$ depends on two field configurations. The expectation values we discussed in the previous sections involved a single field configuration. So we have to adjust our formalism to capture two field configurations. The proposal on how to do this is layed out in \cite{Anninos:2011kh}, which is motivated from spin-glass theory which we briefly review in appendix \ref{app:overlap distribution}. 

In the spin-glass proposal, the expectation value of a quantity that depends on two types of  configurations relies on the probability density for each configuration type. For a field configuration on de Sitter the probability density is given by the amplitude square of the wavefunction, $|\Psi(\Phi)|^2$,
and the spin-glass proposal can be translated into \cite{Anninos:2011kh}
\begin{equation}
\label{appeqn:dSexpect}\langle\mathcal{O}(\Phi_1,\Phi_2)\rangle\equiv\frac{\int\mathcal{D}\Phi_1\mathcal{D}\Phi_2|\Psi(\phi_1)|^2|\Psi(\Phi_2)|^2\mathcal{O}(\Phi_1,\Phi_2)}{\int\mathcal{D}\Phi_1\mathcal{D}\Phi_2|\Psi(\phi_1)|^2|\Psi(\Phi_2)|^2}.
\end{equation}

In computing $\langle d_{12}\rangle$, from among the terms in \eqref{eqn:distance def} the mixed ones in the form of $\Phi^{(1)}\Phi^{(2)}$ do not contribute due to the Gaussian nature of the wavefunction. So $\langle d_{12}\rangle$ boils down to 
\begin{equation}
    \langle d_{12}\rangle=\Bigg\langle\frac{1}{L^d}\int d^dx \left(\Phi^{(1)}(\vec{x})\right)^2\Bigg\rangle+\Bigg\langle\frac{1}{L^d}\int d^dx \left(\Phi^{(2)}(\vec{x})\right)^2\Bigg\rangle.
\end{equation}
Going over to momentum space and ordering the integrals conveniently, we arrive at
\begin{equation}
    \langle d_{12}\rangle=\frac{1}{L^d}\left(\Bigg\langle\int \frac{d^dk_1}{(2\pi)^d} \left(\Phi^{(1)}_{\vec{k}_1}\Phi^{(1)}_{-\vec{k}_1}\right)\Bigg\rangle+\Bigg\langle\int \frac{d^dk_2}{(2\pi)^d} \left(\Phi^{(2)}_{\vec{k}_2}\Phi^{(2)}_{-\vec{k}_2}\right)\Bigg\rangle\right).
\end{equation}
Switching the order of $k$-integration and averaging we have
\begin{align}
    \label{eqn:d12step2}\langle d_{12}\rangle&=\frac{1}{L^d}\Bigg[\int \frac{d^dk_1}{(2\pi)^d}\langle\Phi^{(1)}_{\vec{k}_1}\Phi^{(1)}_{-\vec{k}_1}\rangle+\int \frac{d^dk_2}{(2\pi)^d}\langle\Phi^{(2)}_{\vec{k}_2}\Phi^{(2)}_{-\vec{k}_2}\rangle\Bigg].\end{align} 
    In \eqref{eqn:d12step2}, we recognize the contribution of momentum space two-point functions from section \ref{sec:Wavefunc review}. They automatically contribute with the momentum configuration that give a nonvanishing result by \eqref{twopoint_form}. So we can drop the dirac delta in \eqref{twopoint_form} and set
    \begin{align}
        \langle\Phi^{(i)}_{\vec{k}_i}\Phi^{(i)}_{-\vec{k}_i}\rangle=\frac{1}{8\beta(\vec{k}_i,\eta)}
    \end{align}
    in each one of the integrals in \eqref{eqn:d12step2}. With all these concerns we arrive at
\begin{align}
    \langle d_{12}\rangle&=\frac{1}{L^d}\Bigg[\int \frac{d^dk_1}{(2\pi)^d}\frac{(2\pi)^d}{8\beta(\vec{k}_1,\eta)}+\int \frac{d^dk_2}{(2\pi)^d}\frac{(2\pi)^d}{8\beta(\vec{k}_2,\eta)}\Bigg]\\
    &=\frac{1}{L^d}\int \frac{d^dk}{(2\pi)^d}\frac{1}{4\beta(\vec{k},\eta)}.
\end{align}
As we are interested in $d=3$, 
\begin{align}
\label{eqn:exp d}    \langle d_{12}\rangle&=\frac{1}{L^3}\int \frac{d^3k}{(2\pi)^3}\frac{1}{4\beta(\vec{k},\eta)}.
\end{align}
Looking at equations \eqref{eqn:beta categoric beh}, the integrand in \eqref{eqn:exp d} diverges at $\vec{k}=0$ for all categories and hence the zero mode is problematic for all categories. In section \ref{subsubsec:The trouble with the zero-mode contribution} we justified the removal of the zero-mode for the principal series category by noting that it is not a solution to the equations of motion from the start.  The justification for the massless discrete series case was that the zero-mode corresponds to the gauge degree of freedom. Here the field space calculation is clearly showing that the zero-mode has no room to contribute.

\subsection{Renormalized distance in field space}
\label{subsec:Renormalized distance in field space}
The definition of distance $d_{12}$ in \eqref{eqn:distance def} is the bare definition. There arise two concerns towards using this bare definition of distance:
\begin{enumerate}
    \item \label{item:anninos concern} The expectation value $\langle d_{12}\rangle$ diverges linearly with \emph{proper time} at late-times \cite{Anninos:2011kh}.     
    \item \label{item: roberts concern} Having a finite and non-zero late-time limit on the \emph{width} of the distribution requires certain corrections depending on mass \cite{Roberts:2012jw}.
    \end{enumerate}

   Concern \ref{item:anninos concern} has been pointed out in \cite{Anninos:2011kh} solely by considering the massless case and the proposed remedy is to consider the \emph{renormalized distance} defined as
   \begin{align}
       \delta_{12}=d_{12}-\langle d_{12}\rangle.
   \end{align}
   For the expectation value $\langle d_{12}\rangle$, at late-times setting $\eta=0$, the behavior of $\beta$ according to \eqref{eqn:beta categoric beh} is such that, it is only for the massless case that the integrand in \eqref{eqn:exp d} does not diverge. However in terms of propertime $\tau$, 
\begin{align}\tau&=s= -\frac{1}{H}ln\eta,\end{align} $\langle d_{12}\rangle$ diverges linearly as $\tau\to \infty$ for all categories. 
Although the renormalized distance $\delta_{12}$ was originaly introduced by studying the massless case of discrete series only, as we have just seen the same concern applies to all categories and one needs a renormalized definition of distance for all categories.

Concern \ref{item: roberts concern} is a concern on the \emph{width} of the probability distribution for renormalized distance, $\delta_{12}$. This concern arises by enlargening the results of \cite{Anninos:2011kh} to include massive fields as well. The observation of \cite{Roberts:2012jw} is that the definition of distance in field space needs to be modified depending on the range of mass that corresponds to $\Delta_-=\frac{d}{4}$, with
\begin{equation}
    \Delta_-=\frac{d}{2}-\sqrt{\frac{d^2}{4}-\frac{m^2}{H^2}}
\end{equation}
with the positive root of the square taken into consideration, and the conclusion is that memory always exists on de Sitter while ultrametricity is lost for $\Delta_-=\frac{d}{4}$. This modification at $\Delta_-=\frac{d}{4}$ is introduced due to requiring the width \emph{to have a finite and non-zero late-time limit}. 

Since concern \ref{item: roberts concern} is a concern on the probability distribution function, we would like to close off this discussion with a closer look at the properties of the probability distribution function for distance in field space for all categories of representations. We handle this in the next subsection.

\subsubsection{Revisiting the distance distribution and its width}
\label{subsubsec:probability distribution function PD}
We have noted that the statistical properties of field space for de Sitter was original constructed in \cite{Anninos:2011kh} analogous to spin-glass theory. A key property that both spin-glass systems and fields on de Sitter share is that not all states satisfy cluster decomposition. We have demonstrated this in physical space via two-point functions in section \ref{sec:clust dec in position space}. This property is carefully taken into account in constructing the probability functional in field space. 

The probability functional $\mathbf{P}[\phi]$, of finding a field in state $\phi$ is considered to be decomposed in terms of states that do satisfy cluster decomposition, with some probability  $p_\alpha$, where $\alpha$ labels the states that do satisfy cluster decomposition
\begin{align}
    \label{eqn:alpha decomp} \mathbf{P}[\phi]=\sum_\alpha p_\alpha \mathbf{P}_\alpha[\phi],~~\text{where} \sum_\alpha p_\alpha=1.
\end{align}
The probability distribution for finding a specific distance $\mathcal{D}$ in field space is formally given by
\begin{align}
    \label{eqn: distance dist formal def} \mathbf{P}(\mathcal{D})\equiv \sum_{\alpha,\beta} p_\alpha p_\beta \delta\left(\mathcal{D}-d_{\alpha\beta}\right)
\end{align}
where the distance between two states $\alpha$ and $\beta$ is calculated by $d_{\alpha\beta}=d\left[\langle\phi\rangle_\alpha,\langle\phi\rangle_\beta\right]$. This is distance as defined in \eqref{eqn:distance def}, between local vacuum expectation  values of $\hat{\phi}$ in states $\alpha$ and $\beta$. Equation \eqref{eqn: distance dist formal def} expresses the probability of finding the distance $d_{\alpha\beta}$ between a configuration in state-$\alpha$ which occurs with probability $p_\alpha$, and a configuration in state-$\beta$ which occurs which probability $p_\beta$ to be equal to a specific value $\mathcal{D}$. Following the review of \cite{Castellani_2005}, we give a derivation of how this formal definition for the probability distribution function arises in spin-glass theory in appendix \ref{app:overlap distribution}, where one can see how cluster decomposition principle is worked in. 

In practice it turns out that one cannot carry out these formal calculations, and instead one consults to a thermodynamic limit. In the application of this procedure to de Sitter, the thermodynamic limit corresponds to taking the late-time limit. Then one is calculating distance distributions between configurations of a pair of  fields taken from the Bunch-Davies vacuum state. With the definition in \eqref{appeqn:dSexpect} for expectation values, the distance distribution is defined as 
\begin{subequations}
    \label{eqn: distance dist dS}
\begin{align}
     \mathbf{P}(D)&=\langle\delta(\mathcal{D}-d_{12})\rangle\\
    &\equiv\int \mathcal{D}\phi^{(1)}\mathcal{D}\phi^{(2)}|\Psi_{HH}\left(\phi^{(1)}\right)|^2|\Psi_{HH}\left(\phi^{(2)}\right)|^2\delta\left(D-d\left[\phi^{(1)},\phi^{(2)}\right]\right),
\end{align}
\end{subequations}
where now the probability of being in a configuration of state-$(1)$ is given by $|\Psi_{HH}\left(\phi^{(1)}\right)|^2$. 

Here we are interested in the probability distribution of distance in field space. From a broader perspective, a probability distribution function, $\mathbf{P}(X)$, is related to a moment generating function $G(s)$ via a Fourier transform
\begin{align}\label{eqn:G def} \mathbf{P}(X)=\frac{1}{2\pi i}\int^{i\infty}_{-i\infty}ds e^{s X}G(s).\end{align}
In some cases the moment generating function is easier to calculate then the probability distribution function itself. This is also the case for us. The moment generating function for the renormalized distance $\delta_{12}$ is computed via\cite{Anninos:2011kh}
\begin{align}
    G(s)=\langle e^{-s\delta_{12}}\rangle.
\end{align}
The moment generating function also encodes all the information about the probability distribution, which can be achieved via the cumulant expansion. The cumulant expansion is a power series expansion of the \emph{cumulant generating function} $K(s)$ which is defined as the logarithm of the moment generating function \cite{kardar}
\begin{align}
    K(s)\equiv \log{G(s)}.
\end{align} The first cumulant, $K_1$ gives the mean, the second cumulant $K_2$ the variance and so on. The $n^{th}$ \emph{cumulant}, $K_n$, corresponds to the $n^{th}$ term in the power series expansion of $K(s)$
\begin{equation}
   K(s)=\sum_{n=1}^{\infty} \kappa_n \frac{s^n}{n!}
= \kappa_1 \frac{s}{1!} + \kappa_2 \frac{s^2}{2!} + \kappa_3 \frac{s^3}{3!} + \cdots
,
\end{equation}
which then implies,
\begin{equation}
    \kappa_n = K^{(n)}(0).
\end{equation} 

Lastly we want to discuss the width of the probability distribution of distance in field space, to clarify why Concern \ref{item: roberts concern} requires a correction at mass $m^2=\frac{3}{16}d^2$, which does not coincide with the border of any of the representation theory categories, as depicted in figure \ref{fig:roberts figure}.
 For this purpose, we want to identify the condition for having a finite and nonzero width. Width is usually defined as the standard deviation, which is $\sigma=\sqrt{Var(\delta)}$. That is, width is the square root of the second cumulant. 

For the distance distribution in field space, the moment generating function, $G(s)$ is computed by using \eqref{appeqn:dSexpect}
\begin{align}
        G(s) &= \langle e^{-s\delta_{12}} \rangle = \int \mathcal{D}\Phi^{(1)} \mathcal{D}\Phi^{(2)}|\Psi(\Phi^{(1)})|^2 |\Psi(\Phi^{(2)})|^2 e^{-s\delta_{12}}\\
        &= e^{\langle sd_{12}\rangle} \int \mathcal{D}\Phi^{(1)}_{\vec{k}} \mathcal{D}\Phi^{(2)}_{\vec{k}} e^{-2\int \frac{d^3 k}{(2\pi)^3} \beta(\vec{k},\eta) |\Phi^{(1)}_{\vec{k}}|^2} e^{-2\int \frac{d^3 k}{(2\pi)^3} \beta(\vec{k},\eta)|\Phi^{(2)}_{\vec{k}}|^2} e^{-sd_{12}}.
\end{align}

We then plug in \eqref{eqn:d12 exps in kspace} and write the sum in the exponential as a product 
\begin{equation}\label{G(s) for us}
    G(s)=e^{\langle s\delta_{12}\rangle} \prod_{\vec{k}}^\prime \mathcal{N}_k \int d^2\Phi^{(1)}_{\vec{k}} d^2\Phi^{(2)}_{\vec{k}} e^{-4\beta(\vec{k},\eta) |\Phi^{(1)}_{\vec{k}}|^2 -4\beta(\vec{k},\eta) |\Phi^{(2)}_{\vec{k}}|^2} e^{-2s|\Phi^{(1)}_{\vec{k}}-\Phi^{(2)}_{\vec{k}}|^2/{L^3}}
\end{equation}
where the product is over unordered pairs of $(\vec{k}, -\vec{k})$ which brings on an overall factor of two. 

In equation \eqref{G(s) for us} we have the following Gaussian integral
\begin{equation}
\int d^2\Phi^{(1)}_{\vec{k}} d^2\Phi^{(2)}_{\vec{k}} e^{-4\beta(\vec{k},\eta) |\Phi^{(1)}_{\vec{k}}|^2 -4\beta(\vec{k},\eta) |\Phi^{(2)}_{\vec{k}}|^2} e^{-2s|\Phi^{(1)}_{\vec{k}}-\Phi^{(2)}_{\vec{k}}|^2/{L^3}}=\int d^2x e^{-\frac{1}{2}x^\dagger\mathcal{A} x}
\end{equation}
which is evaluated to be
\begin{align}
    \int d^2\Phi^{(1)}_{\vec{k}} d^2\Phi^{(2)}_{\vec{k}} e^{-4\beta(\vec{k},\eta) |\Phi^{(1)}_{\vec{k}}|^2 -4\beta(\vec{k},\eta) |\Phi^{(2)}_{\vec{k}}|^2} e^{-2s|\Phi^{(1)}_{\vec{k}}-\Phi^{(2)}_{\vec{k}}|^2/{L^3}}=\frac{1}{1+\frac{s}{{L^3}\beta(\vec{k},\eta)}},
\end{align}
with the identification
\begin{align}
    \mathcal{A}=4\begin{bmatrix}
        \frac{s}{{L^3}}+\beta(\vec{k},\eta)&-\frac{s}{{L^3}}\\
        \frac{s}{{L^3}}&\frac{s}{{L^3}}+\beta(\vec{k},\eta)
    \end{bmatrix},~~x=\begin{bmatrix}\Phi^{(1)}_{\vec{k}}\\ \Phi^{(2)}_{\vec{k}}
    \end{bmatrix}.
\end{align}
Taking into account the overall normalization, we reach the expression of \cite{Anninos:2011kh} for the moment generating function
\begin{equation}
    G(s) = e^{\langle s\delta_{12}\rangle} \prod_{\vec{k}}^\prime \frac{1}{1+\frac{s}{{L^3}\beta(\vec{k},\eta)}} \\
    = \prod_{\vec{k}}^\prime \frac{e^{s/{L^3}\beta(\vec{k},\eta)}}{1+\frac{s}{{L^3}\beta(\vec{k},\eta)}}.
\end{equation}
From here the cumulant generating function K(s), for the distance distribution in field space becomes 
\begin{align}\label{cgf}
    K(s)&=\log(G(s))=\log(\langle e^{-s\delta_{12}}\rangle) \\
    &=\sum_{\vec{k}}^\prime \left( \frac{s}{{L^3}\beta(\vec{k},\eta)}-log(1+\frac{s}{{L^3}\beta(\vec{k},\eta)})\right).
\end{align}

For small cumulants (small $s/\beta(\vec{k},\eta)$), we can expand the logarithm term in \eqref{cgf} as  
\begin{equation}\label{log:powerexp}
    \log(1+\frac{s}{{L^3}\beta(\vec{k},\eta)}) = \sum_{n=1}^\infty \frac{(-1)^{n+1}}{n}\left(\frac{s}{{L^3}\beta(\vec{k},\eta)}\right)^n.
\end{equation}
Thus we have the power series expansion of the moment generating function $G(s)$, which gives us the cumulant generating function $K(s)$ as follows
\begin{align}
    K(s)&=\sum_{n=2}^{\infty} \frac{(-1)^n s^n}{n} \sum_{\vec{k}}^\prime \frac{1}{\left({L^3}\beta(\vec{k},\eta)\right)^n} \\
    &= \sum_{n=2}^{\infty} \frac{(-1)^n s^n}{n{L^{3n}}} \int_{0}^\infty \frac{d^3k}{\beta(\vec{k},\eta)^n}
\end{align}
The sum above starts from $n=2$ because the first term in the expansion \eqref{log:powerexp} is canceled out by the first term in \eqref{cgf}. The variance is 
\begin{equation}\label{eqn:var}
Var(\delta_{12})=\kappa_2={\frac{4\pi}{L^6}}\int_{0}^\infty \frac{dk}{\beta(\vec{k},\eta)^2}.
\end{equation}
Looking at equations \eqref{eqn:beta categoric beh}, $\beta(\vec{k},\eta)$ has complicated $k$-dependence, making the integration in \eqref{eqn:var}, which involves $\frac{1}{\beta^2}$ difficult. However, to get an overall idea we can consider the superhorizon limit.

In the superhorizon limit, we noted that all categories exhibit similar behaviour as
\begin{equation}
    \beta_{k|\eta|\ll1}\sim\frac{k^{d-2\Delta_-}}{|\eta_0|^{2\Delta_-}}.
\end{equation}
So for $d=3$, in the superhorizon limit we can approximate the variance as
\begin{align}
    \label{eqn:var superhor}
\kappa_2^{k|\eta|\ll1}&=\frac{4\pi|\eta_0|^{4\Delta_-}}{{L^6}}\int_{k_{IR}}^{k_{UV}\ll\frac{1}{|\eta_0|}} \frac{dk}{k^{6-4\Delta_-}}\\
&=-\frac{4\pi|\eta_0|^{4\Delta_-}}{{L^6}}\left(6-4\Delta_-\right)\left[\frac{1}{k^{6-4\Delta_-}}\right]^{k_{UV}\ll\frac{1}{|\eta_0|}}_{k_{IR}}
\end{align}
with a UV and IR cutoff. Under the superhorizon approximation the variance only vanishes for $\Delta_-=\frac{3}{2}$ which occurs for the case of the massless scalar.
\vspace*{1cm}

\section{Conclusions and outlook}
\label{sec:Conclusions and outlook}

In this work we gathered some clues towards locality on de Sitter with respect to unitary irreducible representations of the de Sitter group. To achieve this we studied free scalar fields of different mass on $dS_4$, both in physical space and in field space with focus on principal and discrete series representations from among the unitary irreducible representations of the de Sitter group in four spacetime dimensions ($SO(4,1)$). The principal series represent heavy matter fields, while discrete series capture gauge theory on de Sitter. Both perspectives relied on the late-time behavior of two-point functions, which we calculated via the de Sitter wavefunction and Euclidean sphere path integral, as reviewed in section \ref{sec:Review of two-point functions on the $S^4$ and $dS_4$}. Our goal was to see how the discussion on locality can be categorized from the point of view of representation theory. We retrieved the familiar result that the principal series representations seem to exhibit locality while the discrete series representations exhibit special properties that violate it. However we gained some new perspectives from a more careful treatment of the zero-modes.

Zero modes require care. In physical space they seem to be problematic only for discrete series representations. This observation relies on analytical continuation to the  sphere, which we review in section \ref{sec:zeromodes} and it can be understood in terms of gauge fixing as we point out in section \ref{subsubsec:The trouble with the zero-mode contribution}. Although the zero-modes do not cause any problems in the principals series sphere two-point function, it still makes sense to remove them as they do not correspond to a solution of the equations of motion. In field space, complementary to the physical space the situation is more clear. Zero modes are explicitly problematic for all types of representations, as we revisit with an explicit analysis in section \ref{subsec:field space problem with the zero-mode}.

In physical space we saw that principal series representations satisfy cluster decomposition via the power law decay of the late-time two-point function in terms of position space separation $|\vec{x}-\vec{y}|$, which we obtained in section \ref{subsec:Principal series two-point function in position space}, while the discrete series late-time two-point function exhibits logarithmic growth as we obtained in section \ref{subsec:Discrete series two-point function in position space}. The logarithmic growth of the discrete series has been pointed out in earlier literature \cite{Anninos:2011kh,Benincasa:2022gtd}. Here, we carry on this discussion further by doing a zero-mode analysis in section \ref{subsubsec:Removing the zero-mode}. We explicitly show that the discrete series logarithmic growth persists even if the zero-mode is removed. Hence this violation of locality is a curious feature of discrete series representations, and it is a physical fact. 

In field space cluster decomposition is explored via the probability distribution in field space. In this analysis the difficulty is formulating a well defined definition of distance. The well defined distance is expected to have an expectation value that does not  diverge at late-times \cite{Anninos:2011kh} and lead to a distribution with a finite width at late-times \cite{Roberts:2012jw}. We have labeled these requirements as Concern \ref{item:anninos concern} and \ref{item: roberts concern} in section \ref{subsec:Renormalized distance in field space}. With an explicit calculation we justify the treatment of Concern \ref{item:anninos concern}. An exact calculation of the variance for principal and discrete series field space probability distribution is not so straight forward. We do a cumulant expansion and focus on the superhorizon limit to get an estimate on the behavior of the variance in section \ref{subsubsec:probability distribution function PD}. The interesting aspect of the superhorizon limit is, all categories show similar behavior as we point out in section \ref{sec:Wavefunc review}, in accordance with \cite{Roberts:2012jw}. However when it comes to the variance the superhorizon limit immediately signals a problem with the discrete series, which lies at one end one the spectrum as depicted in figure \ref{fig:roberts figure}. In the superhorizon limit the variance vanishes only for the discrete series case. While the analysis of \cite{Roberts:2012jw} is more involved,  interestingly, the superhorizon limit makes a distinction with respect to the representation categories.

The analysis of locality on de Sitter, in terms of cluster decomposition, shows interesting and counterintuitive aspects even at the level of scalar fields. Matter and gauge fields seem to be treated differently when it comes to cluster decomposition. It will be interesting to broaden this conclusion by considering bosons of nonzero spin and fermions in future work.

\acknowledgments

It is with great pleasure that we thank Dionysios Anninos, Tarek Anous, Atsushi Higuchi, Vasileios Letsios, Ben Pethybridge and Tonguç Rador for insightful discussions and to Paolo Benincasa for his very helpful comments on an earlier version of this manuscript. Both MÖ and GŞ acknowledge support from TÜBİTAK 2232 - B International Fellowship for Early Stage Researchers program with project number 121C138. GŞ also acknowledges support from Boğaziçi University Start Up Grant no 19858 in the final stages of this work.

\appendix
\section{Review of the overlap distribution}
\label{app:overlap distribution}
In this appendix, following \cite{Castellani_2005}, we review a derivation of equation \eqref{eqn: distance dist formal def} for the probability distribution for finding a specific distance $\mathcal{D}$ in field space. 

First let's consider how cluster decomposition effects computing expectation values. Consider an observable quantity $\mathcal{O}$ made up of configurations $\sigma$. For instance in a spin glass system $\sigma$ can denote the value of each individual spin, a bunch of which make up a state $\alpha$, such that $\sigma\in\alpha$. If the Hamiltonian of this system is denoted as $H(\Sigma)$, where the system can be divided into a number of states $\alpha$, $\beta$, etc with configurations labeled by $i,j$ and $\Sigma$ involves interactions of all such states, the partition function is
\begin{align}
    \label{appeqn:partition function}Z=\int \mathcal{D}\sigma e^{\beta H(\Sigma)}.
\end{align}
The expectation value of an observable quantity $\mathcal{O}$, made up of states of configurations of the system can be calculated by
\begin{align}
    \label{appeqn:expectation value def} \langle\mathcal{O}\rangle=\frac{1}{Z}\int \mathcal{D}\sigma e^{\beta H(\Sigma)}\mathcal{O}(\sigma).
\end{align}
To explore the effects of cluster decomposition, we are interested in the expectation value of
\begin{align}
    \mathcal{O}=\sigma_i\sigma_j
\end{align}
where the two configurations belong to different states, $i\in \alpha$, $j\in\beta$ and the expectation value is calculated by
\begin{align}
   \label{appeqn av sigmasigma} \langle\sigma_i\sigma_j\rangle=\frac{1}{Z}\sum_\alpha\int_{i\in \alpha}\mathcal{D}\sigma_i\sum_\beta\int_{j\in \beta}\mathcal{D}\sigma_j e^{\beta H(\Sigma)}\sigma_i\sigma_j.
\end{align} 
If these two states are physically far apart, that is in the limit $|i-j|\to \infty$, by cluster decomposition we expect that there is no interaction between the two states and hence between the configurations coming from these two states, in which case we can factorize the Hamiltonian and the partition function as
\begin{align}
H(\Sigma)=H(\sigma_i)+H(\sigma_j),~~Z=Z_\alpha Z_\beta~~\text{as}~~|i-j|\to \infty,
\end{align}
where
\begin{align}
Z_\gamma=\int_{i\in\gamma}\mathcal{D}\sigma_ie^{\beta H(\sigma_i)},
\end{align}
and the probability $p_\gamma$ of having state $\gamma$ is
\begin{align}
\label{appeqn:weight}p_\gamma=\frac{Z_\gamma}{Z}.
\end{align}
Then
\begin{align}
\langle\sigma_i\rangle=\frac{1}{Z_\alpha}\sum_\alpha \int_{i\in\alpha}\mathcal{D}\sigma_i e^{\beta H(\sigma_i)}\sigma_i.
\end{align}
In the limit $|i-j|\to\infty$, the integrals in \eqref{appeqn av sigmasigma} are unrelated and can be evaluated individually, leading to
\begin{align}
    \label{appeqn:clustering prop}\langle\sigma_i\sigma_j\rangle=\langle\sigma_i\rangle\langle\sigma_j\rangle,~~\text{as}~~|i-j|\to\infty.
\end{align}
This is called the clustering property. And we see that the quantum field theory definition of cluster decomposition really leads to a statistical cluster decomposition.

Going back to our main quest, we want to calculate the probability distribution of finding the overlap between two states $q_{\alpha\beta}$ to be a specific value $Q$. First let's consider the overlap $q_{\alpha\beta}$ itself.

 The overlap $q_{\alpha\beta}$ is a measure of the similarity between states and is evaluated by considering the configuration $\sigma_i$ existing in different states as \cite{Castellani_2005}
\begin{align}
\label{appeqn:overlap_states}    q_{\alpha\beta}=\frac{1}{N}\sum^N_{i=1}\langle\sigma_i\rangle_\alpha\langle\sigma_i\rangle_\beta
\end{align}
where $N$ is the number of configurations that make up the state. One can also consider the similarity between configurations $\sigma$ and $\tau$ coming from different states $\sigma\in\alpha$ and $\tau\in\beta$, with the same number of configurations \cite{Castellani_2005}
\begin{align}
\label{appeqn:overlap_config}    q_{\sigma\tau}=\frac{1}{N}\sum^N_{i=1}\sigma_i\tau_i.
\end{align}
The overlap of states is related to the overlap of configurations. Starting by expanding out the expectation values in equation \eqref{appeqn:overlap_states}
\begin{align}
    q_{\alpha\beta}=\frac{1}{N}\sum^N_{i=1}\frac{1}{Z_\alpha}\int_{\sigma\in\alpha}\mathcal{D}\sigma\sigma_i e^{-\beta H(\sigma)}\frac{1}{Z_\beta}\int_{\tau\in\beta}\mathcal{D}\tau\tau_i e^{-\beta H(\tau)},
\end{align}
and changing the order of summation and integration we arrive at
\begin{align}
    \nonumber q_{\alpha\beta}&=\frac{1}{Z_\alpha Z_\beta}\int_{\sigma\in\alpha}\mathcal{D}\sigma\int_{\tau\in\beta}\mathcal{D}\tau e^{-\beta H(\sigma)}e^{-\beta H(\tau)}\frac{1}{N}\sum^N_{i=1}\sigma_i \tau_i \\
    &=\frac{1}{Z_\alpha Z_\beta}\int_{\sigma\in\alpha}\mathcal{D}\sigma\int_{\tau\in\beta}\mathcal{D}\tau e^{-\beta H(\sigma)}e^{-\beta H(\tau)}~q_{\sigma\tau}
\end{align}
as given in \cite{Castellani_2005}.

In the thermodynamic limit, the expectation value of the overlap of configurations $q_{\sigma\tau}$ to be a particular value $Q$ is expected to be calculated via 
\begin{align}
    P(Q)&=\langle\delta(Q-q_{\sigma\tau})\rangle\\
    &=\frac{1}{Z^2}\int\mathcal{D}\sigma\mathcal{D}\tau e^{-\beta H(\sigma)-\beta H(\tau)}\delta(Q-q_{\sigma\tau}).
\end{align}

Consider our configurations to belong to different states, $\sigma\in\alpha$, $\tau\in\beta$ and using \eqref{appeqn:weight} rewrite the above expression as follows
\begin{align}
    P(Q)&=\langle\delta(Q-q_{\sigma\tau})\rangle\\
   \label{appeqn:PQcalc} &=\sum_{\alpha,\beta}p_\alpha p_\beta \frac{1}{Z_\alpha}\int_{\sigma\in\alpha}\mathcal{D}\sigma\frac{1}{Z_\beta}\int_{\tau\in\beta}\mathcal{D}\tau e^{-\beta H(\sigma)-\beta H(\tau)}\delta(Q-q_{\sigma\tau}).
\end{align}
Considering the definition of $q_{\sigma\tau}$ in equation \eqref{appeqn:overlap_config} this expression can be thought of as 
\begin{align}
P(Q)&=\sum_{\alpha,\beta}p_\alpha p_\beta \frac{1}{Z_\alpha}\int_{\sigma\in\alpha}\mathcal{D}\sigma\frac{1}{Z_\beta}\int_{\tau\in\beta}\mathcal{D}\tau e^{-\beta H(\sigma)-\beta H(\tau)}\delta(Q-\frac{1}{N}\sum_{i}\sigma_i\tau_i)\\
    &=\sum_{\alpha,\beta}p_\alpha p_\beta \delta\left(Q-\frac{1}{N}\sum_i\langle\sigma_i\tau_i\rangle\right).
\end{align}
In the presence of cluster decomposition, if the configurations $\sigma$ and $\tau$ are taken far away from each other, we have
\begin{align}
P(Q)&=\sum_{\alpha,\beta}p_\alpha p_\beta \delta\left(Q-\frac{1}{N}\sum_i\langle\sigma_i\rangle_\alpha\langle\tau_i\rangle_\beta\right)\\
&=\sum_{\alpha,\beta}p_\alpha p_\beta \delta\left(Q-q_{\alpha\beta}\right),
\end{align}
which gives the expectation value of the overlap of configurations taking on a particular value $Q$ in terms of the expectation value of the overlap of states taking on that particular value. This is the starting point of equation \eqref{eqn: distance dist formal def}. One may not be able to perform the weighted sum but one still has the hope of calculating an expectation value which is what the right hand side of equation \eqref{eqn: distance dist dS} is. When working on de Sitter instead of working with spin glasses one has the wavefunction doing the job of the partition function and the overlaps $q_{\alpha\beta}$ are understood as distances in field space $d_{12}$.

\bibliography{refs}
\bibliographystyle{JHEP}
\end{document}